\documentclass[12pt,a4paper,titlepage]{article}

\usepackage{t1enc}
\usepackage{amsmath}
\usepackage{latexsym}
\usepackage{natbib}
\usepackage{epsfig}
\usepackage{textcomp}
\usepackage{alltt}
\usepackage{graphicx}
\usepackage{lscape}
 \usepackage{hyperref}
 
\usepackage{multirow}
\usepackage{numprint}
 \usepackage{xcolor}
  
\usepackage{rotating}
\usepackage{pdflscape}

\usepackage{setspace}
\usepackage{booktabs}

\usepackage{amsfonts}

\usepackage{titlesec}
\titleformat*{\section}{\large\bfseries}
\titleformat*{\subsection}{\normalsize\bfseries}

\usepackage[nolists, tablesfirst, nomarkers]{endfloat}
\DeclareDelayedFloatFlavor{sidewaystable}{table}

\begin{document}

\setlength{\abovedisplayskip}{5 pt}
\setlength{\belowdisplayskip}{5 pt}

\newcommand{\bi}{\begin{itemize}}
\newcommand{\ei}{\end{itemize}}
\newcommand{\nothere}[1]{}
\newcommand{\noi}{\noindent}
\newcommand{\mbf}[1]{\mbox{\boldmath $#1$}}
\newcommand{\cond}{\, |\,}
\newcommand{\hO}[2]{{\cal O}_{#1}^{#2}}
\newcommand{\hF}[2]{{\cal F}_{#1}^{#2}}

\newcommand{\tl}[1]{\tilde{\lambda}_{#1}^T}
\newcommand{\la}[2]{\lambda_{#1}^T(Z^{#2})}
\newcommand{\I}[1]{1_{(#1)}}
\newcommand{\cd}{\mbox{$\stackrel{\mbox{\tiny{\cal D}}}{\rightarrow}$}}
\newcommand{\cp}{\mbox{$\stackrel{\mbox{\tiny{p}}}{\rightarrow}$}}
\newcommand{\cas}{\mbox{$\stackrel{\mbox{\tiny{a.s.}}}{\rightarrow}$}}
\newcommand{\ld}{\mbox{$\; \stackrel{\mbox{\tiny{def}}}{=3D} \; $}}
\newcommand{\nk}{\mbox{$n \rightarrow \infty$}}
\newcommand{\con}{\mbox{$\rightarrow $}}
\newcommand{\dprime}{\mbox{$\prime \vspace{-1 mm} \prime$}}
\newcommand{\Borel}{\mbox{${\cal B}$}}
\newcommand{\bevis}{\mbox{$\underline{\em{Proof}}$}}
\newcommand{\Rd}[1]{\mbox{${\Re^{#1}}$}}
\newcommand{\il}[1]{{\int_{0}^{#1}}}
\newcommand{\pl}[1]{\mbox{\bf {\LARGE #1}}}

\newtheorem{thm}{Theorem}
\newtheorem{lemma}{Lemma}
\newtheorem{prop}{Proposition}

\pagestyle{empty}

\begin{center}
{\Large{\bf Efficient estimation of regression models for recurrent events in presence of terminal events}}
\end{center}
\vspace{1 cm}

\begin{center}
Giuliana Cortese  \\
Department of Statistical Sciences, University of Padova \\
Via Cesare Battisti 241, 35121 Padova, Italy \\
gcortese@stat.unipd.it

\vspace{4mm}
 and \\  
 
 \vspace{4mm}
Thomas H. Scheike\\
Section of Biostatistics, Department of Public Health, University of Copenhagen \\
\O ster Farimagsgade 5, DK-1014 Copenhagen N, Denmark\\
ts@biostat.ku.dk\\ \vspace{4mm}
\end{center}

\bigskip
\centerline{\sc Abstract}

Recurrent events are common in biomedical studies, and efficient methods are needed to analyse them, particularly when a terminal event may occur. Regression modelling of the marginal mean of recurrent events provides a simply interpretable approach, but standard estimation based on inverse probability of censoring weighting may not fully exploit observed event histories in multiple-events settings. We demonstrate how the popular Ghosh-Lin proportional means model can be estimated efficiently by developing an augmentation procedure that is also applicable to composite endpoints. Using semiparametric efficiency theory, we characterize the fully efficient estimator. First, we investigate its construction under frailty models. We then propose a dynamic censoring-augmented estimator that is straightforward to compute and does not require a fully specified joint model for recurrent and terminal events. We show theoretically that it improves on the standard Ghosh–Lin estimator and derive an explicit expression for its asymptotic variance reduction. We also establish conditions under which tests of no treatment effect remain valid in randomized trials when the proportional means model is misspecified. Extensive simulations examine performance under dependence between recurrent and terminal events, unobserved heterogeneity, censoring-model misspecification, and auxiliary covariates. The practical gain of the proposed methods is illustrated with an analysis of catheter-related bloodstream infections in patients with chronic intestinal failure receiving home parenteral support.

\bigskip

\noi
\vspace{3mm}

\noi

\newpage
\pagestyle{plain}
\setcounter{page}{1}
\section{Introduction}

The use of recurrent events modelling has gained increased popularity in recent years both in theoretical work 
and in practical settings. In the pharmaceutical field, a recent industry qualification, see \cite{akacha2018request},
advocated that going beyond a time to first event analysis can be of interest because it more fully describes 
the phenomenon studied and often it will also lead to a more efficient use of the data. This is particularly important
in the setting of expensive randomized clinical studies. 
On the theoretical side, new developments are ongoing and a recent summary can be found 
in \cite{cook-lawless-2007}, see also \cite{andersen-ravn-2023,cook-lawless-2018}. 

The regression modelling of recurrent events can roughly be grouped into intensity-based or rate-based 
modelling where the intensity/rate of the events of interest can be modelled alongside
the intensity/rate of the terminal event. This line of work includes, for example, the Andersen-Gill model \citep{ag-82}, the Prentice-Williams-Peterson model
\citep{pwp-1981}, and the  Wei-Lin-Weissfeld and  Lin-Wei-Yang-Ying models \citep{wei:lin:weissfeld:1989,lin:wei:yang:ying:2000}.

Another line of work instead has  been focussing on what is called marginal mean regression
modelling, which aims to model the marginal mean of the recurrent events. Here notably, the Ghosh-Lin model has been a pioneering approach that has been frequently used in applications \citep{Ghosh2002}. There have been several extensions of the Ghosh-Lin model to combine 
multiple events into composite endpoints \citep{mao-lin-2016} and, along the same lines,
making composite endpoints in the context of competing risks \citep{furberg-leader-2022}.
The Ghosh-Lin model is a Cox-type regression  extension of the simple nonparametric 
marginal mean estimator that was considered by \cite{cook-lawless-1997,ghosh-lin-2000} and
recently, it was shown how the marginal mean parameter can be estimated efficiently \citep{cortese-scheike-2022}. 

The Ghosh-Lin model and its composite endpoint extensions are semi-parametric Cox-type models
and the estimated parameters and inference for these models are often based on inverse probability 
of censoring (IPCW) adjustment. 
The current paper presents how this estimation procedure 
 can be easily made more efficient by building on 
the theory of efficient estimation for censored data; see, e.g., 
\citet{tsiatis2006semiparametric,van2003unified,robins-rotnitzky:1992,bang:tsiatis:2000} 
In addition, we show that the efficiency gain can be quite large and leads to an important advantage: it further introduces
certain double robustness properties that make the parameter estimation less sensitive to the IPCW model selection for censoring, which is typically needed to provide unbiased estimates.
We show theoretically how to obtain the most efficient Ghosh-Lin type estimator that relies on 
correctly specifying and estimating the working models involved in the augmented term of the estimating equations. 
Therefore, we propose a novel approach that develops augmented estimating equations based on a dynamic prediction model with time-varying coefficients for the augmented term due to censoring.
The performance of this estimator is studied in extensive realistic simulations and the novel method gives a surprisingly large efficiency gain. The suggested estimator has the advantage of being computationally 
quite simple and keeping the computation time linearly in the sample size of the data. 

We use our developments in the context of studying catheter-related bloodstream infections (CRBSI) 
in patients with chronic intestinal failure receiving human parenteral nutrition (HPN); see \citet{Tribler2018}. 
These data are  characterized by strong heterogeneity between patients, 
because of high variability both in the number of CRBSI and in the risk of death, 
and possible correlation between recurrent events and death. 
Recurrent regression models have found important use in  many other applications, such as, e.g., in the LEADER randomized controlled trial to study the effect of treatment on cardiovascular recurrent events \citep{marso}.

\section{Notation and regression models}
\label{sect2}

Let $N^{*}(t)$ be the counting process of the number of recurrent events observed over a time-period $[0,t]$, where $t\leq\tau$ for some constant $\tau$, and $D$ denote the survival time to a terminal event. If a subject has the terminal event,  he/she can not experience further recurrent events after $D$. Thus we only observe the recurrent event processes up to $\tau \wedge D$, where
$a \wedge b = \min(a,b)$, such that  $N^*(t)=N^{*}(t \wedge D)$.

Our interest is in studying the effect of a $p$-dimensional covariate vector $X$ measured at baseline on the marginal mean number of recurrent 
events over time.
For this scope, 
\cite{Ghosh2002} suggested 
the semiparametric multiplicative regression model 
\begin{align}
	E(N^*(t) | X)  & = \mu_0(t) \exp(\beta^T X ) , \label{eqr:1}
\end{align}
where $\mu_0(t)$ is the baseline marginal mean number of recurrent events for $X=0$ and $\beta$ gives the covariate effects.
The proportional means model in \eqref{eqr:1} 
has the advantage that it does not require any assumptions on
dependence among recurrent events, and dependence between recurrent and terminal events,
and has the scope to quantify the covariate effects marginally. 
 Unlike Poisson processes, such models are very general and flexible, and accommodate for heterogeneity between subjects, who can experience no, few or many recurrent events, as well as a terminal event. We assume no
dependence structures among recurrent events, but the scope is here to recover
such information directly from the history of the recurrent event process over
time using a dynamic increment estimation based on semiparametric efficiency
theory for missing data. 

In the most general case, it is enough to assume that both the recurrent events
process and the survival distribution of $D$ depend on covariates, and thus
obtain an indirect regression model for the marginal mean: 
$$
   \mu(t |X)=E(N^*(t) |X)=\int_0^t S(s |X) d R(s |X),
$$
where $S(t|X)=P(D> t |X)$ and the conditional rate $dR(t) = E(dN^*(t) | D> t,
X)$ describes the recurrent event rate among the survivors. Note that, since no
further recurrent events can be experienced after the terminal event, then for
$t> D$ it is $N^*(t) = N^*(D)$,  and thus $E\{dN^*(t) |X \} = E\{dN^*(t \wedge
D)|X ) = E(dN^* (t) I(D \geq t) |X \}$. Unfortunately, this indirect modeling
approach is not able to measure covariate effects directly on the marginal
mean of recurrent events, which is here our primary scope.

The observation of recurrent and terminal events may be precluded for subjects
who are right censored during follow-up, and thus the process $N^*(t)$ is only
observable up to the censoring time $C$. The standard assumption of conditional
independent censoring is considered here, i.e., that $C$ is independent of $D$
and $N^*(t)$ conditional on $X$. Let us define $\delta=I(D \leq C)$ and $T=D
\wedge C$. Moreover, let $N(t)=N^*(t \wedge C)$ be the observed number of
events and define the at-risk process $Y(t)=I(T\geq t)$. 
We consider a random sample of size $n$ and assume that data come from the independent observations  $\{N_{i}(t), T_i , \delta_i, X_i \}$
 for $i=1,...,n$ and any $t \in [0,\tau]$.

\subsection{Inference and asymptotic results}

In presence of right-censoring, \cite{Ghosh2002}  suggested estimating the parameters of the proportional mean model in \eqref{eqr:1} by inverse probability of
censoring weighted (IPCW) score equations as follows. For simplicity, let us consider the case where the censoring distribution does not
depend on covariates. 
Let $G(t)$ be the survival distribution of $C$ and $\hat G(t)$ be its Kaplan-Meier estimator.  
In presence of a terminal event, either $C_i$ or $D_i$ is observed.
Accordingly, let us define the IPC weights as
$w_i(t)= I(C_i \geq D_i \wedge t) G(t) / G(T_i \wedge t)$. These weights can be shown to have mean value $E( w_i(t)) = E[ E\{  w_i(t) | D_i \}] = G(t) E[ E\{ I(C_i \geq D_i \wedge t) / G(D_i \wedge t) |D_i \} ] = G(t)$ for $i=1, \ldots, n$, by the law of conditional expectations, and they can be estimated by the observed counterpart $\hat w_i(t) = I(C_i \geq D_i \wedge t) \hat G(t) / \hat G(T_i \wedge t)$. Then, an estimate $\hat \beta$ of the regression coefficients is obtained as a solution to the weighted estimating equation 
\begin{align}
\label{eqr:4}
 U(\beta) = \sum_{i=1}^{n} \int_0^{\tau} \hat w_i(t) \{ X_i - \bar{X}(\beta, t) \} d N_i(t) = 0,              
\end{align}
where $\bar{X}(\beta, t) = \hat S_1(\beta,t) / \hat S_0(\beta,t)$ 
with 
$\hat S_k(\beta, t) = \sum_{i=1}^n  \hat w_i(t)  X_i^k \exp{(X_i^T\beta)}$, for $k=0,1,2$. 
The estimator $  \hat \mu_0(t)$ of the baseline marginal mean function is consistent and has asymptotically a  zero-mean Gaussian distribution. 
See  Ghosh \& Lin (2003), Ghosh (2000) and Lin et al (2003) for additional details and proofs on asymptotic properties.
If neither the terminal event nor the right censoring is present in the data, i.e., the data are fully observed, the weights simplify 
such that $\hat w_i(t) =\tilde w_i(t) = 1$.

Let us define the quantity
$  A =  E [ \partial U_i (\beta_0) / \partial \beta ] = E [ \int_0^{\tau} (X_i - \bar x(\beta_0,t))^{\otimes 2} G(t) e^{X_i^T \beta_0} d \mu_0(t) ] $, with $ \bar x(\beta,t) = s_1(\beta,t) / s_0(\beta,t)$, where $s_k(\beta,t) = E[G(t)X^{\otimes k} e^{X^T \beta } ] $ is the limit of $\hat S_k (\beta,t)$ for $k=0,1,2$,
and $ \hat A(\beta) = - n^{-1} \sum_{i=1}^n \partial U_i(\beta) / \partial \beta$.
It was shown that $\hat \beta$ is a consistent estimator of $\beta_0$ and that  the $i$-th
influence function  of $(\hat \beta - \beta_0 )$ is $A^{-1} \phi_i(X,\beta_0) $, where $\phi_i(X,\beta_0)   = \phi_i^R(X,\beta_0) + \phi_i^C(\beta_0)$
and 
\begin{align}
	\phi_i^R(X,\beta_0) = &  \int_0^{\tau} \{ X_i - \bar x(\beta_0, t) \} w_i(t) dM_i(t,X),  \label{eq1:phiR} \\
\phi_i^{C}(\beta_0) = &  \int_0^{\tau} \frac{ q(t)}{S(t) G(t)} d M_i^C(t),
\label{eq1:phiC}
\end{align}
with $S(t)$ being the survival function of $D$, 
and
\begin{align}
 q(t) & = - E \left[ I(T_i < t) \, \int_t^{\tau}  \{X_i - \bar x(\beta_0, s) \}  \, w_i(s) \, dM_i(s,X) \right]  \nonumber \\
      & = - E \left[ I(D_i < t) \ \frac{I(D < C)}{G(D)} \, \int_t^{\tau}  \{X_i - \bar x(\beta_0, s) \} G(s) \, dM_i(s,X) \right] \nonumber  \\
      & = - E \left[  I(D_i < t) \, \int_t^{\tau}  \{X_i - \bar x(\beta_0, s) \} G(s) \, dM_i(s,X) \right] .
      \label{eq:qt}
\end{align}
Equation \eqref{eq:qt} involves the basic mean-zero process $M_i(s,X) = N_i(t) - \mu_0(t) \exp(X^T \beta)$ and the censoring martingale
 $M^C_i(t)= N^C_i(t) -  \int_0^t Y_i(s) d\Lambda^C (s)$, 
based on the  counting process $N^C_i(t) = I(T_i \leq t, \delta =0)$ and its cumulative hazard  $\Lambda^C (t) = - \log G(t) $. 
Appendix A proves the decomposition given in \eqref{eq1:phiR} and \eqref{eq1:phiC}, in line with \citet{Ghosh2002}.

In addition, it was shown that the normalized estimator $n^{1/2}(\hat \beta - \beta_0 )$ is asymtotically distributed as $N(0, \Sigma)$, with $\Sigma= A^{-1} \text{Var}\left\{\phi_1(X,\beta_0)\right\} (A^{-1})^T$ and $\text{Var} \{\phi_1(X,\beta_0) \} = 
E \{ [\phi^R_1(X,\beta_0) + \phi^C_1(\beta_0) ]^{\otimes  2} \}$, 
where the outer product is $y^{\otimes  2}= y\, y^T$.

\section{IPCW estimation}
\label{IPCW}

In this section, by applying  
the standard IPCW technique in Robins \& Rotnitzky (1992), we derive an IPCW estimating equation and the related variance, which will then be used to derive the final efficient estimators in the next section.

First, 
weights are rewritten as 
$w_i(t) = G(t) \, r_i(t) $ where the following identity holds:
\begin{align*}
	r_i(t) =  \frac{I(C_i \geq D_i \wedge t)}{ G(T_i \wedge t)} =  1- \int_0^{t}    \frac{ 1}{ G(u)} Y_i(u)   dM_i^C (u)  
\end{align*}
and therefore, replacing this identity in  $ \phi_i^R(X,\beta_0)$ and changing order of integration, we obtain that
\begin{align}
 \phi_i^R(X,\beta_0) = &  \int_0^{\tau} \{ X_i - \bar x(\beta_0, t) \} ~ \frac{I(C_i > D_i \wedge t) }{G(D_i \wedge t)} ~ G(t) ~ dM_i(t,X),  \\
	   = & ~ H_i(0,X) - \int_0^\tau  H_i(t,X) \frac{ 1}{ G(t)} Y_i(t) ~  dM_i^C (t)  \label{eq2:phiR}
\end{align}
with $H_i(t,X) = \int_t^\tau \{ X_i - \bar x(\beta_0, s) \}  ~ G(s) ~ dM_i(s,X)$.

In addition, when the Ghosh-Lin model is correct, it follows that $E(H(t,X)) = 0$ for all
$t$, and then $E(H(t,X)) = E(H(t,X) I(D<t)) + E(H(t,X) I(D>t))=0$. 
Thus, by the law of conditional expectations, we can also rewrite the term $q(t)$ in \eqref{eq:qt} as
a conditional mean given that the subject is still alive with no terminal event,
\begin{align*}
	q(t)  &  =  E \left[ H_i(t,X) I(D_i > t) \right] = S(t) E \left[ H_i(t,X) | \, D_i > t \right] 
\end{align*}
and obtain that 
\begin{align}
\label{eq2:phiC}
	\phi_i^{C}(\beta_0) = & \int_0^{\tau} E \left[ H_i(t,X) | \, D_i > t \right] \frac{ 1}{ G(t)} ~ Y_i(t) ~ d M_i^C(t).
\end{align}

Therefore, we can  write the $i$-th influence function, ignoring the scaling with $A$, as the sum of a full-data score, $H_i(0,X)$, and 
an orthogonal censoring term, as follows:
\begin{align}
	\phi_i(X)  &  = H_i(0,X)  - \int_0^\tau \left[ H_i(t,X) - E(H_i(t,X) | D > t ) \right] \frac{ 1}{G(t)} ~ Y_i(t) ~ dM_i^C (t).
    \label{infl_func}
\end{align}

As a consequence of this alternative IPCW formulation, the  variance of the i-th influence function $\phi_i(X)$ can also be derived as sum of two components:
\begin{align}
	E \left[ \phi_i(X)^{\otimes 2}\right]  &  = E \left[   H_i(0,X)^{\otimes  2}\right]  + 
	E\left[   \int_0^\tau \{ 
 \left[ H_i(t,X) - E(H_i(t,X) | D > t ) \right] I(D_i>t) \}^{\otimes 2} \frac{ 1}{G(t)} d \Lambda^C(t) \right],
	\label{var-1}
\end{align}
The second term of $\phi_i(t)$ is still a martingale 
\citep{bang:tsiatis:2000} and thus has conditional mean 
zero, given $D$ and $N(t)$. 
Therefore, the above result is obtained by computing directly the variance of the censoring martingale and by using the conditional independence of the censoring given covariates.
Note that the contribution of the Kaplan–Meier estimator of $G(t)$ to the influence function is the extra term containing $E(H_i(t,X) | D > t )$ in the censoring martingale integral. Thus,  the terms $H_i(t,X)$ are centered to their conditional mean and as a consequence of this result, the variance of the influence function, and thus the variance of the estimating equation for $\beta$, is reduced, as compared with the variance computed when the true $G(t)$ is used.

\vspace{0.5 cm}
\noindent {\it {\bf Remark}:}  
A useful extension is to allow the censoring distribution to depend on covariates via strata 
$L(X) \in \{ 1, \ldots, K\}$, defined from the covariates $X$, such that 
$G(t,L(X)) = P( C > t | L(X)) = P(C>t | X)$. The resulting weights are 
$w_i(t,L(X)) = G_c(t,L(X)) I(C > D \wedge t) / G_c(D \wedge t, L(X))$, where the censoring distribution can be estimated by the stratified Kaplan-Meier estimator. This approach leads to  the influence functions
 \begin{align*}
	\phi_i^R(X,\beta_0) = &  \int_0^{\tau} \{ X_i - \bar x(\beta_0, t) \} w_i(t,L(X)) dM_i(t,X),  \\
\phi_i^{C}(\beta_0) = &  \int_0^{\tau} \frac{ q(t,L(X))}{S(t,L(X)) G(t,L(X))} d M_i^C(t,L(X)),
\end{align*}
where $S(t,L(X)) = P( D > t | L(X))$,  $H_i(t,X) =  \int_t^{\tau} \{ X_i - \bar x(\beta_0, s) \} G(s,L(X)) dM_i(s,X)$, and
$q(t,L(X)) =  - E [ H_i(t,X) I(D_i < t) | L(X) ]$, and with the censoring martingale now depending on $L(X)$.

Since  $E( H(t,X) | L(X)) = E( E[ H(t,X) | X, L(X)] | L(X)) = 0$ under the Ghosh-Lin model where $E(H(t,X) | X)= 0$, 
it is still true that 
 \begin{align*}
q(t, L(X)) & = - E( H(t,X) I(D < t ) | L(X)) =  E( H(t,X)  I( D> t)  | L(X)), \\
           & = S(t,L(X)) E( H(t,X) | D> t, L(X)).
\end{align*}
Therefore,  an important result is achieved: The variance of the influence function $E \left[ \phi_i(X)^{\otimes 2}\right]$ is further reduced since, in equation \eqref{var-1}, the second component $ E[ \{ [ H(t,X) - E( H(t,X) | D>t, L(X))] I(D>t) \}^{\otimes 2} ]$ is smaller than or equal to 
$ E[ \{ [ H(t,X) - E(H(t,X) | D>t)] I(D>t) \}^{\otimes 2}]$.

\section{Efficiency from censoring augmentation}

Based on results from Section \ref{IPCW}, we  show how the IPCW Ghosh-Lin estimator can be made fully efficient, by finding the optimal censoring augmentation term 
as suggested by \citet{tsiatis2006semiparametric,van2003unified,robins-rotnitzky:1992}. 
This  requires adapting the standard theory to 
 the specific structure of the Ghosh-Lin estimator, which can then yield an appropriate efficiently augmented estimating equation.

To derive the fully efficient Ghosh-Lin estimator for $\beta$, an augmented inverse probability weighted complete-case estimating function is needed. This function is easily obtained from the non-scaled observed-data influence function 
$\phi_i(X,\beta_0)   = \phi_i^R(X,\beta_0) + \phi_i^C(\beta_0)$, with
$\phi_i^R(X,\beta_0) $ given as in \eqref{eq2:phiR}  and $\phi_i^C(\beta_0)$ written as in \eqref{eq2:phiC}.
In the latter, the conditional mean therein, $E \left[ H_i(t,X) | \, D_i > t \right]$, is the key function that allows the full efficiency to be achieved. Indeed, by applying semiparametric efficiency theory for coarsened data, this function, when assumed to be known, provides the optimal censoring augmentation term, and thus the most efficient estimator for $\beta$, with the smallest asymptotic variance. 

The variance of the estimator for $\beta$ is equivalent to the variance of its influence function. 
From equation \eqref{var-1}, we note that the variance of the non-scaled influence function of the Ghosh-Lin estimating equation (ignoring the scaling with the limit of the second derivative), is reduced thanks to the orthogonal censoring term therein.
Then, we have  directly observed that the optimal censoring augmentation can be obtained by adding 
\begin{align*}
	&   \int_0^\tau E [ H_i(t,X) |  {\cal H}_i(t), D_i > t ]  \frac{1}{G(t)} Y_i(t) dM_i^C (t) 
\end{align*}
to the Ghosh-Lin estimating equation, 
because the censoring term of the variance in \eqref{var-1} will then be minimized. The notation ${\cal H}_i(t)$ represents the individual observed history up to time $t$ that is assumed to be predictable \citep{abgk-book}. 

The efficiently augmented estimating equation thus becomes
 \begin{align}
U(\beta)^A = & \sum_{i=1}^{n} \int_0^{\tau} \hat w_i(t) \{ X_i - \bar{X}(\beta, t) \} d N_i(t)  
	  +  \sum_{i=1}^{n} \int_0^{\tau} E[ H_i(t,X)  | {\cal H}_i(t), D_i >t  ] \frac{1}{ \hat G(t)} d \hat M_i^C(t)
\label{eq:aug}
 \end{align}
where 
the first term involves contributions for the time period of complete observation of recurrent events data, and the second term includes additional contributions that recover the missing information due to right censoring. The second term, denoted as the augmentation term, depends on the conditional mean of $H_i(t,X)$ given the 
history ${\cal H}_i(t)$ of the $i^{th}$ subject up to time $t$ among those that have not experienced the terminal event yet, i.e.,
those with $D_i >t$. The history includes the cumulative number of recurrent events of subject $i$ up to time $t$ and when they took place.

 \subsection{Augmentation based on random effects models}
 
Computing in practice the augmentation term requires complete modelling of the recurrent events process, $N^*(t)$, as well as the terminal event 
$D$ and their dependence. For obtaining individual predictions of the conditional mean $E[ H_i(t,X)  | {\cal H}_i(t), D_i >t  ]$ needed in the augmentation, a possible approach is to utilize various random effects models in relation to the type of dependence structure in the data.  To further increase the statistical
efficiency, these models can also be allowed to depend on time-constant covariates $X$.

Hereafter, we consider three scenarios where regression models are based on different types of dependencies. Define $\alpha_d(t|X) = \alpha_{0d}(t)\exp{(X^T \beta_d)}$ and $\alpha_1 (t|X)= \alpha_{01}(t) \exp{(X^T \beta_1)}$. The nonparametric functions $\alpha_{0d}(t)$ and  $\alpha_{01}(t)$ represent the baseline rates for an individual with $X=0$.

\textbf{Model 0:} The terminal event and the recurrent
events are completely independent, and heterogeneity is only due to covariates $X$.
A multiplicative regression model $\lambda_d(t|X) = \alpha_d(t|X)$ 
is assumed for the terminal event rate and a similar model 
$\lambda_1(t|X) =\alpha_1(t|X)$ for the recurrent event rate, such that $D$ is independent of $N^*(t)$ and $N^*(t)$ has independent increments. 
In this case, for any $s$ and $t$, with $s>t$, it is
 \begin{align*}
E[dN_{d,i} (s) | {\cal H}_i(t), D_i>t ]  & =\alpha_{d}(s|X_i) ds, \\
E [ dN^*_i(s) | {\cal H}_i(t), D_i > t ] & = E [ dN^*_i(s) | D_i > t ] = \alpha_1(s|X_i) ds. 
 \end{align*}
Then, with this model, the augmentation term is not null, depends on $X_i$ but not on $N_i(t-)$.

\textbf{Model 1:}  The terminal event is independent of the recurrent events, but some additional heterogeneity is introduced by allowing possible association between recurrent events.
This model imposes that $\lambda_d(t|X) = \alpha_{d}(t|X)$ 
and $\lambda_1(t |Z, X) = Z \, \alpha_1 (t|X)$, where the random effect $Z$ has a Gamma distribution with $E(Z)=1$ and $Var(Z) = \theta$.  
Thus, the marginal recurrent event rate for a survivor at $t$, for any $s: s>t$, is
 \begin{align*}
& E[dN_{d,i} (s) | {\cal H}_i(t), D_i>t ]   =\alpha_{d}(s|X_i) ds,  \\
& E[dN_i (s) | {\cal H}_i(t), D_i>t ]  
=  E [  Z_i | {\cal H}_i(t), D_i > t] \, \alpha_{r}(s|X_i) ds =  \frac{ 1+\theta N_i (t-)}{1+ \theta A_{01}(t)\exp{(X_i^T \beta_1)}}  \alpha_1 (s|X_i) ds, 
 \end{align*}
where $A_{01}(t) =\int_0^t \alpha_{01}(u)du$. 

\textbf{Model 2:}  Possible dependence involves both the terminal event and the recurrent events, and thus, an even larger heterogeneity is introduced. 
In this model, the terminal event and recurrent event rates are assumed to be
$\lambda_d(t|Z,X) = Z \alpha_{d}(t|X)$ 
and $\lambda_1 (t |Z, X) = Z \, \alpha_1 (t|X)$, with a common random effect Z distributed as in Model 1. In this case, the marginal  terminal event and recurrent event rates for a survivor at $t$ are, respectively, for any $s: s>t$, of the form
\begin{align*}
E[dN_{d,i} (s) | {\cal H}_i(t), D_i>t ]  &
= \frac{ 1+\theta N_i (t-)} {1+ \theta \{ A_{01}(t)\exp{(X_i^T \beta_1)} + A_{0d}(t)\exp{(X_i^T \beta_d)} \}} \,\alpha_{d}(s|X_i) ds , \\
E[dN_i (s) | {\cal H}_i(t), D_i>t ]  &
= \frac{ 1+\theta  N_i (t-)}{ 1+ \theta \{A_{01}(t)\exp{(X_i^T \beta_1)} + A_{0d}(t)\exp{(X_i^T \beta_d)} \}} \,  \alpha_1 (s|X_i) ds,
 \end{align*}
where $A_{0d}(t) =\int_0^t \alpha_{0d}(u)du$.

To compute the conditional mean $E[ H_i(t,X)  | {\cal H}_i(t), D_i >t  ]$ in the augmentation under any of the considered Models 0, 1 and 2,   it is convenient to decompose the  function $H_i(t,X)$ as follows:
\begin{align*}
H_i(t,X) & = \int_t^\tau \{ X_i - \bar x(\beta_0, s) \} \, G(s)  \, dM_i(s,X)  \\
 & = \int_t^\tau I(D_i > s) \{ X_i - \bar x(\beta_0, s) \} \, G(s) \, dN_i(s) - 
 \int_t^\tau  \{ X_i - \bar x(\beta_0, s) \} \, G(s) \, d\mu_0(s) \exp{(X_i^T \beta)} \\
 & = H_i^N(t,X) - H_i^{\mu}(t,X), 
\end{align*}
since in the first integrand, it is $dN_i(s)=0$ for $s \in (D_i, \tau]$.  
Given that the component $H_i^{\mu}(t,X)$ is a function only of $s$ and $X_i$ and does not depend on $N$ and $D$, and thus it is constant with respect to the conditional mean, we obtain
\begin{align}
\label{H}
\begin{split}
E[ H_i(t,X) | {\cal H}_i(t), D_i >t ] & =  \int_t^\tau \frac{S(s | {\cal H}_i(t))}{S(t| {\cal H}_i(t))} \{ X_i - \bar x(\beta_0, s) \} \, G(s)  \, E[dN_i (s) | {\cal H}_i(t), D_i>t ]  
- H_i^{\mu}(t,X)
\end{split}
\end{align}
where, for any $s$ s.t. $s>t$, it is
$ S(s | {\cal H}_i(t))/ S(t| {\cal H}_i(t)) 
= \exp{ \{- \int_t^s E[dN_{d,i} (u) | {\cal H}_i(t), D_i>t ]  \} }$.
The conditional mean in \eqref{H} is an explicit function of the marginal recurrent event rate $E[dN_i (s) | {\cal H}_i(t), D_i>t ]$ and the terminal event rate $E[dN_{d,i} (u) | {\cal H}_i(t), D_i>t ]$. Therefore, by imposing any of the previously described random effects regression models (Models 0, 1, and 2) for these rates, the conditional mean can be computed for each subject and for any $t$. 

Individual predictions of the above conditional mean are not easily obtained, because estimates of all regression coefficients and involved quantities are required, and this process is difficult to perform for each $t$ spanning $[0,\tau]$. Moreover, the optimal efficiency will not be reached since the individual history information can not be fully captured; finally, the random effect (frailty) modeling approach requires knowing the assumptions on the dependence structure among recurrent events and the terminal event. 

Therefore, to solve the above-mentioned issues related to random effects models, in the next section we suggest an alternative approach that provides improved efficiency by estimating the conditional expectation via a 
 dynamic regression augmentation that extends the regression augmentation approach of \cite{bang:tsiatis:2000}. 
 In this flexible approach, the strong parametric dependence assumptions of random effect models are relaxed, and time-varying regression coefficients are considered to model the effect of the recurrent events and covariates history. 

 To build the novel approach, we note that, when random effects Models 0, 1, and 2 are assumed, the different functional forms of the conditional mean in \eqref{H} can yield very useful suggestions about the possible structure of the dynamic regression models to be imposed in the augmentation term, where $N(t-)$ and $X$ may enter as linear or nonlinear functions.  Therefore, applying some algebraic manipulations, we obtain that

\textbf{Model 0:} $E[ H_i^N(t,X)  | {\cal H}_i(t), D_i >t  ] \equiv \gamma_{00}(X_i, t)$ and thus the conditional mean is independent of the recurrent process history $N_i(t-)$; the resulting model can be summarized as a function only of covariates, with a generic functional form $\gamma_{00}(X_i, t) = \int_t^{\tau} e^{f_{00}(X_i,s)} f_{01}(X_i, s)ds$, being $f_{00}$ and $f_{01}$ some non specified functions of $X_i$.
    
\textbf{Model 1:} $E[ H_i^N(t,X)  | {\cal H}_i(t), D_i >t  ] \equiv \gamma_{01}(X_i, t) + \theta \gamma_{01}(X_i, t) \, N_i(t-)$, i.e, a simple linear transformation of $N(t-)$; the functional form in $X_i$ has a more complex structure since we obtain  $\gamma_{01}(X_i, t)= \gamma_{00}(X_i, t) / (1+ \theta A_{01}(t)e^{X_i^T \beta_1} )$
   . Note that, when $\theta=0$, this model reduces to Model 0.
   
\textbf{Model 2:} We obtain linear transformations of $N(t-)$ that have coefficients being time varying with $s$, and this leads to a more complex structure of the conditional mean, given by
   $E[ H_i^N(t,X)  | {\cal H}_i(t), D_i >t  ] \equiv \int_t^{\tau} 
   e^{\gamma_{02}(X_i, s) + \, \theta \, \gamma_{02}(X_i, s) \, N(t-)} \, [\gamma_{12}(X_i, s) + \, \theta \, \gamma_{12}(X_i, s) \, N(t-) ] ds $. Functions $\gamma_{02}$ and $\gamma_{12}$ have here a complex nonlinear structure of $X_i$ that is difficult to write explicitly.

\subsection{Estimation based on dynamic-regression augmentation}

The idea of the dynamic-regression censoring augmentation 
is to replace 
the optimal prediction on the full space of the history, $E[ H_i(t,X)  | {\cal H}_i(t), D_i >t  ]$, given within equation \eqref{eq:aug}, with a linear
combination of $K$ specified regressors $e(t)^T = (e_1(t),....,e_K(t))$, and thus, 
regress the conditional mean on the space spanned by $e(t)$, 
as in  \cite{bang:tsiatis:2000,cortese-scheike-2022}.  This approach leads to the optimal augmentation based on 
dynamic regression with time-varying coefficients $\gamma(t) = (\gamma_1(t), \ldots, \gamma_p(t))$, where each component is a $K$-dimensional vector given by
 \begin{align}
	 \gamma_j(t) & =   E [ (e(t) - \bar e(t))^{\otimes 2} ]^{-1}  E [ H_{ij}(t,X) \{e(t) -\bar e(t) \}  Y(t) ],
     \qquad j=1, \ldots, p,
  \label{eq:timecoeff}
 \end{align}
 with $H_{ij}(t,X)$ being the $j$-th element of $H_{i}(t,X)$.
The quantity $\bar e(t) = \sum_i Y_i(t) e_i(t) / \sum_i Y_i (t)$ is the at-risk average of the subject-specific predictors $e_i(t)^\top = (e_{i1}(t),\ldots, e_{iK}(t))$.
Note that, if the Ghosh-Lin regression model for the mean number of recurrent events is based on $p$ covariates, the dynamic regression approach is applied to each component of the $p$-dimensional augmentation term in the estimating equation \eqref{eq:aug}.
Then, we would augment with 
$\gamma_j(t)^T (e_i(t) - \bar e(t)) Y_i(t)$ instead of $E[ H_{ij}(t,X)  | {\cal H}_i(t), D_i >t  ]$, for $j=1, \ldots, p$,
where $e_i(t)$ is the vector of regressors observed over the $i$-th individual.
We note that, for example,  when $N(t-)$ is used among the regressors $e(t)$, then in equation \eqref{eq:timecoeff} the inverse will typically not exist for the initial times, 
due to all values being zero from the start;  therefore, a generalized inverse can be used in practice to deal with this issue. 
Another useful observation is that $\gamma(0)=0$ since $E(H(0,X,\beta_0))=0$ by construction and also $\gamma(\tau)=0$. 
The typical functional form 
f the  coefficients $\gamma_{jk}(t)$, for $k=1,\ldots, K$,
is thus a U-shaped curve, reflecting the importance of covariates on the prediction of the 
different components of the score. This form will also indicate that coefficients are time-varying, and thus non-constant over the time period $[0,\tau]$.

 Under this dynamic censoring-augmented approach, the efficient estimating equation is 
 \begin{align*}
U^L(\beta) = & \sum_{i=1}^{n} \int_0^{\tau} \hat w_i(t) \{ X_i - \bar{X}(\beta, t) \} d N_i(t)  \\
	 & +  \sum_{i=1}^{n} \int_0^{\tau} \left\{ \gamma(t)^T ( e_i(t) - \bar e(t))  \right\} \frac{1}{ \widehat G(t)} d \hat M_i^C(t),
 \end{align*}
and given $b_i(t) = \left\{ \gamma(t)^T ( e_i(t) - \bar e(t))  \right\} \frac{1}{ \widehat G(t)}$, the augmentation term can be equivalently written as 
$ \sum_{i=1}^{n} \int_0^{\tau} b_i(t) d \widehat N_i^C(t) = \sum_{i=1}^{n} I( C_i < D_i) I(T_i \leq \tau) \, b_i(T_i)$, since the compensators in $\hat M_i^C(t)$ cancel out after summing over subjects. This last formulation shows that only predictions for censored observations contribute to the augmentation term; however, all observations are needed for the at-risk average $\bar e(t)$ and for estimating  $\gamma(t)$.

The efficient influence function related to equation \eqref{eq:aug} results in
\begin{align*}
\phi_i^L(X) = \phi_i(X) +  \int_0^{\tau} \left\{ \gamma(t)^T [ e_i(t) - E(e_i(t) \mid D_i >t) ] \right\} \frac{1}{G(t)} d M_i^C(t),
\end{align*} 
with $\phi_i(X)$ given as in equation \eqref{infl_func}. Due to the additional term in $\phi_i^L(X)$,
its variance results to be $Var\{\phi_1^L(X) \} = Var\{\phi_1(X)\} - V(\gamma)$, 
where $Var\{\phi_1(X)\}$ is given in equation \eqref{var-1}.  
The variance reduction obtained by using the dynamic regression augmentation, as compared to the simple Ghosh-Lin estimator, is quantified by the term  
\begin{align*}
	V(\gamma) &	= E \left[ \int_0^\tau \gamma(t)^T \{ e(t)-\bar{e}(t)\}^{\otimes 2} \gamma(t) \, \frac{Y(t)}{ G^2(t)} d \Lambda^C(t) \right].
\end{align*}
This term can be quite substantial and the variance reduction is obtained (asymptotically) no matter what regression model has been used for the augmentation, thus leading us to reach an efficiency gain in any case, e.g., in case of model mispecification. The degree of increase in efficiency depends 
on how well the $K$ specified regressors $e(t)$ 
the tail of the score for the Ghosh-Lin model.

For the dynamic augmentation, we also explored regression models where the regression coefficients have a parametric time-structure of the form $\gamma(t) = \gamma \cdot f(t,\theta)$; among the possible functions $f(t)$ to be chosen, a simple example is the parabola 
 $f(t)=(t \cdot (t-\tau))$, which was also used for the simulation studies reported in the next section.
 With this dynamic parametric-time regression, the optimal regression
 parameter is constant and equal to  
 \begin{align}
   \tilde  \gamma = E\left\{ \int_0^\tau f(t)^2 [ e(t)-\bar{e}(t)]^{\otimes 2} \frac{Y(t)}{ G^2(t)} d \Lambda^C(t) \right\}^{-1} 
	              E \left\{ \int_0^\tau  H(t,X) f(t) ( e(t)-\bar{e}(t)) \frac{Y(t)}{ G^2(t)} d \Lambda^C(t) \right\}.
                  \label{gamma-const}
 \end{align}
 When $f(t) \equiv 1$ as in \citet{bang:tsiatis:2000}, and thus the regression coefficients in the augmentation model are time constant, we found that no gain was obtained in our extensive 
simulations, which are shown later, but for the parabola shape we obtained a small gain. 

 To estimate the optimal dynamic regression coefficients in \eqref{eq:timecoeff} that minimize the variance of the estimating equation, we must replace $H(t,X)$ with its IPCW estimator. We thus obtain
 \begin{align*}
 \hat \gamma(t) & = \left\{ \sum_i [ e_i(t)-\bar{e}(t)]^{\otimes 2} Y_i(t) \right\}^{-1} \hat G(t) \sum_i  \widehat H_i^{ipcw}(t,X) \, [ e_i(t)-\bar{e}(t)] \, Y_i(t),
 \end{align*}
 with $H_i^{ipcw}(t, X)= \int_t^{\tau} \{X_i - \bar x(\beta_0, s) \} w_i(s) d M(s,X_i)$, and
 with $\widehat H_i^{ipcw}$ indicating the corresponding quantity with the use of the estimated censoring weights $\hat w_i(s)$.
 The derivation of the estimator $ \hat \gamma(t)$ is based on the property that  
$E\{ H(t,X) [ e(t)-\bar{e}(t)] Y(t) \} = G(t) E\{ H(t,X) [ e(t)-\bar{e}(t)] I(D>t) \}$ that equals 
$G(t)  E\{ H^{ipcw}(t,X) [ e(t)-\bar{e}(t)] I(D>t) \} ) = G(t) E\{ H^{ipcw}(t,X) [ e(t)-\bar{e}(t)] Y(t) \}$, using also that 
the covariate vector $e(t)$ is defined from the baseline covariates and the history of $N$.

\vspace{0.5 cm}
\noindent {\it {\bf Remark}:}  
An important and useful property of the dynamic censoring augmentation is that this will always reduce variance, even when the 
Ghosh-Lin model is not valid. Regressing $H(s,X)$ on $Y(s) ( e(s) - \bar e(s))$ leads to the same variance reduction as when regressing $H(s,X)+q(t)$. This is in contrast to using a larger censoring model that only leads to a guaranteed variance reduction when the model
holds and $-q(s,L(X))$ equals $E[ H_i(s,X) \mid D \geq s, L(X)] \, S(s,L(X))$.

\section{The randomized clinical trial setting}

We finally state a robustness result that holds for the null hypothesis of no treatment effect, 
even when using a miss-specified Ghosh-Lin model in the context of a study with randomized clinical trials (RCT).
We consider covariates $Z=(A,X)$  where $A$ is a binary treatment indicator, and $X$ are additional 
covariates. In the RCT setting, in addition, $A$ and $X$ are independent due to the randomization. 
Assume that  $\lambda(t; A,X)$ is  the true marginal mean model $E(dN(t) | A,X)$, and that we
use a working model given in Ghosh-Lin form 
with covariates $Z$.  
It follows, using the arguments of \cite{stru:kalb:1986}, that
$\hat \beta$, the optimizer of the Ghosh-Lin estimating equation,
will converge towards the unique $\beta^*$ (under weak regularity conditions) that solves the equation
\begin{align}
    \int_0^\tau \left[ s_1(t) - \frac{s_1(t,\beta)}{s_0(t,\beta)} s_0(t) \right] dt 
    \label{eq:robGL}
\end{align}
with  $s_j(t) = E( Z^j \lambda(t;A,X) w(t))$ and $s_j(t,\beta) = E( Z^j \exp(Z^T \beta) w(t))$, for $j=0,1$.

Under the independent censoring assumption, 
if there is no effect of treatment on the marginal mean
given covariates, such that $E(N(t) | A,X)$ does not depend on $A$, then the Ghosh-Lin working model 
that satisfies condition 1 or 2, as follows, 
\begin{enumerate}
    \item using a correctly specified censoring model, $G(t;A,X)$, possibly, depending on covariates, and then using
    stabilized weights on the form 
    $w(t) = G(t, A) I(C > D \wedge t)/G(D \wedge t; A, X)$ 
    for fitting the Ghosh-Lin model where $G(t,A)$ 
    is some function that only depends on $(t,A)$;
    \item or, if the terminal event does not depend on $A$ 
    and if the censoring has an additive hazards form $\lambda_c(t,A,X)=\lambda_{c,0}(t)+A \lambda_{c,1}(t) + \lambda_{c,2}(t,X)$, 
       then  using a censoring adjustment working model based on an additive hazards model, for example,
    $\lambda_c^{W^*}(t,A,X)=\tilde \lambda_0(t)+A \tilde \lambda_1(t)$, a censoring model stratified on $A$;
\end{enumerate}
will lead to correct identification of no treatment effect, in the sense that $\beta^*_1=0$. 
Therefore, if robust standard errors are used, then we will also get correct type-I error in RCT's. 
The important point is not to generate any interaction terms between $X$ and $A$.
Both conditions are a direct consequence of 
computing the simple $s_j$, and the second condition relies on the
fact that, with a true additive hazards model, then the applied additive censoring model will correctly adjust 
for $\lambda_1(t) A$, extending the arguments of \cite{Blanche2022} to this setting, 
in the computation of  $E(w(t) | A, X)$.   Various other censoring working models may be used as long as they
do not introduce interaction/correlation between $A$ and $X$. We prove the above robustness result in Appendix B.

\section{Simulations}
\label{sim}

Let $\lambda_d(t)$ denote the terminal event rate and $\lambda_1(t)$ the recurrent event rate. 
To investigate the performance and efficiency of the proposed estimators, we simulated data from three regression models with different dependence structures among the recurrent events $N^*(t)$ and the terminal event $D$. Model 0 assumes that $D$ is independent of $N^*(t)$ and $N^*(t)$ has independent increments: $\lambda_d(t) = \alpha_d(t) \exp(X^T \beta_d)$ and $\lambda_r(t) =    \alpha_1(t)\exp(X^T \beta_1)$. Model 1 also imposes that $D$ is independent of $N^*(t)$, but accounts for association between recurrent events by incorporating unobserved heterogeneity: $\lambda_d(t) =  \alpha_d(t)\exp(X^T \beta_d)$  and $\lambda_1(t) = Z \, \alpha_1(t)\exp(X^T \beta_1)$, with frailty $Z$ being a Gamma random variable with mean 1 and variance $\theta$. Model 2 assumes a possible dependence between the terminal event and recurrent events, and also among recurrent events: $\lambda_d(t) = Z  \,\alpha_d(t)\exp(X^T \beta_d)$ and $\lambda_1(t) =Z \, \alpha_1(t)\exp(X^T \beta_1)$, with the same $Z$.  
In Models 0,1,2, we considered a binary variable $X_1$ having Binomial distribution with $P(X_1=1)=0.5$, and a continuous variable $X_2$ with Standard Normal distribution. 
The regression coefficients in the terminal event and recurrent events rate models were set to, respectively, $\beta_d=(0.3,-0.3)$ and $\beta_1=(0.3,-0.3)$. 
For the assumed Models 0,1 2, there was only minor variation across the number of expected recurrent events and the size of the random effect.

\begin{figure}
    \centering
    \includegraphics[width=.8\linewidth]{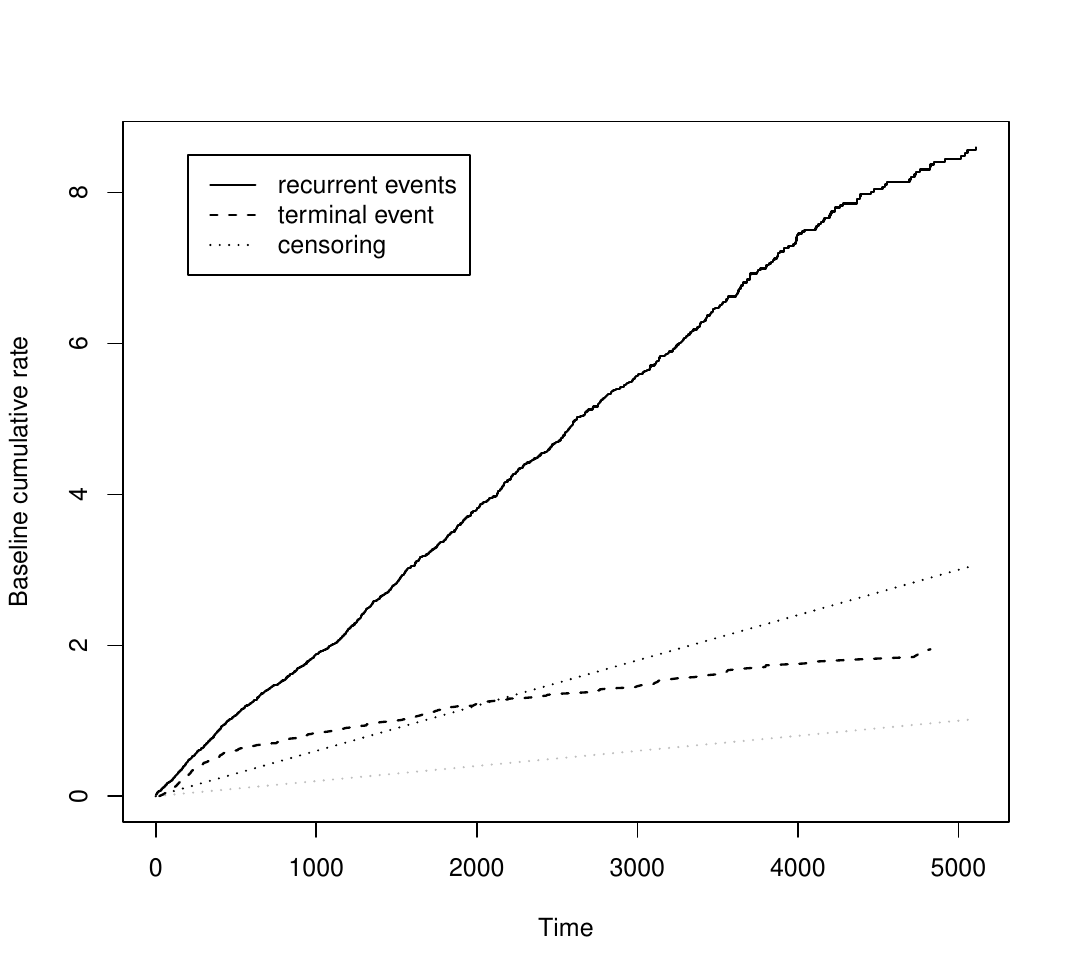}
    \caption{Baseline cumulative rates for the recurrent events and terminal event (from the real-data application on CBRSI), and cumulative hazard of exponential censoring times ($k_c=1$: gray, $k_c=3$: black), used for data simulation.}
    \label{fig:cumrate_sim}
\end{figure}

We simulated data that resemble the real data used for the application in the next section, thus letting  $\alpha_d(t)$ and $\alpha_1(t)$
be piecewise linear approximations of the terminal event rate 
and the recurrent events rate that we observed among 
survivors in the data, see Figure \ref{fig:cumrate_sim}.

This leads to a baseline survival rate of around $14$ \% at 5000 days,
and an average baseline mean number of recurrent events of 
around $2.3$ at $5000$ days. 
The censoring time was exponentially
distributed with hazard $\lambda_c=k_c/5000$ and 
different levels of censoring $k_c=1,3$,
the highest level thus leading to a censoring proportion of 
about $81$\% at 5000 days. 
We also considered different levels of dependence via the variance of the random effect $Z$, by assuming $\theta \in \{0.5, 1\}$. 
We generated 5000 independent replications of each simulation setting with 
sample sizes $n=200,400,800$. 

We then fitted a standard Ghosh-Lin model (denoted \textit{`base'}) with covariates $X_1$ and $X_2$, and a Ghosh-Lin model  (denoted \textit{`cens'}) where the censoring distribution in IPCW estimation was allowed to depend on four strata given by the interaction between the binary covariates $X_1$ and $\tilde X_2$, the latter obtained by dichotomizing the continuous covariate $X_2$ around its median. 
In addition, we fitted an augmented estimator of the standard Ghosh-Lin model, with three different dynamic regression models for prediction of $E[ H(t,X)  | {\cal H}_i(t), D_i >t  ]$ in the censoring augmentation term: 
Model \(m\)-A1 used the predictor vector $e(t)= (N(t-),X_1, X_2)$ corresponding to 3 regression coefficients; model \(m\)-A2 additionally included an interaction term, with $e(t)= (N(t-),X_1, X_2, X_1*X_2)$ and 4 regression coefficients; the last more complex model \(m\)-A3 included  $e(t)=( N(t-), X_1, X_2, \tilde X_2, X_1*X_2, N(t-)* X_1, N(t-)*\tilde X_2, X_1* \tilde X_2, N(t-)* X_1* \tilde X_2)$ and 9 coefficients.
In the prediction models \(m\)-A1, \(m\)-A2, \(m\)-A3, the time-varying coefficients $\gamma(t)$ were assumed to have either a nonparametric functional form, as given in equation \eqref{eq:timecoeff} (\textit{Dynamic Regression}), or a parametric structure  $\gamma(t) = \gamma \cdot f (t, \theta)$ with parabolic function $f (t, \theta)= t \cdot (t - \tau)$ and $\gamma$ given in equation \eqref{gamma-const} (\textit{Regression-$f(t,\theta$)}).
The regression augmented estimators were based on computing the
augmentation term one time, and then refitting the model. 

Table \ref{glmean} shows the mean of the estimated regression coefficients $\beta$ shown for all models, only for
$n=200$, as there were no major differences for the other sample sizes. We note
that the means are consistent across all the different models but depend on the
amount of censoring, since the models used for data generation do not lead to a mean of recurrent events
that is on Ghosh-Lin form.

\begin{figure}
    \centering
    \includegraphics[width=1\linewidth]{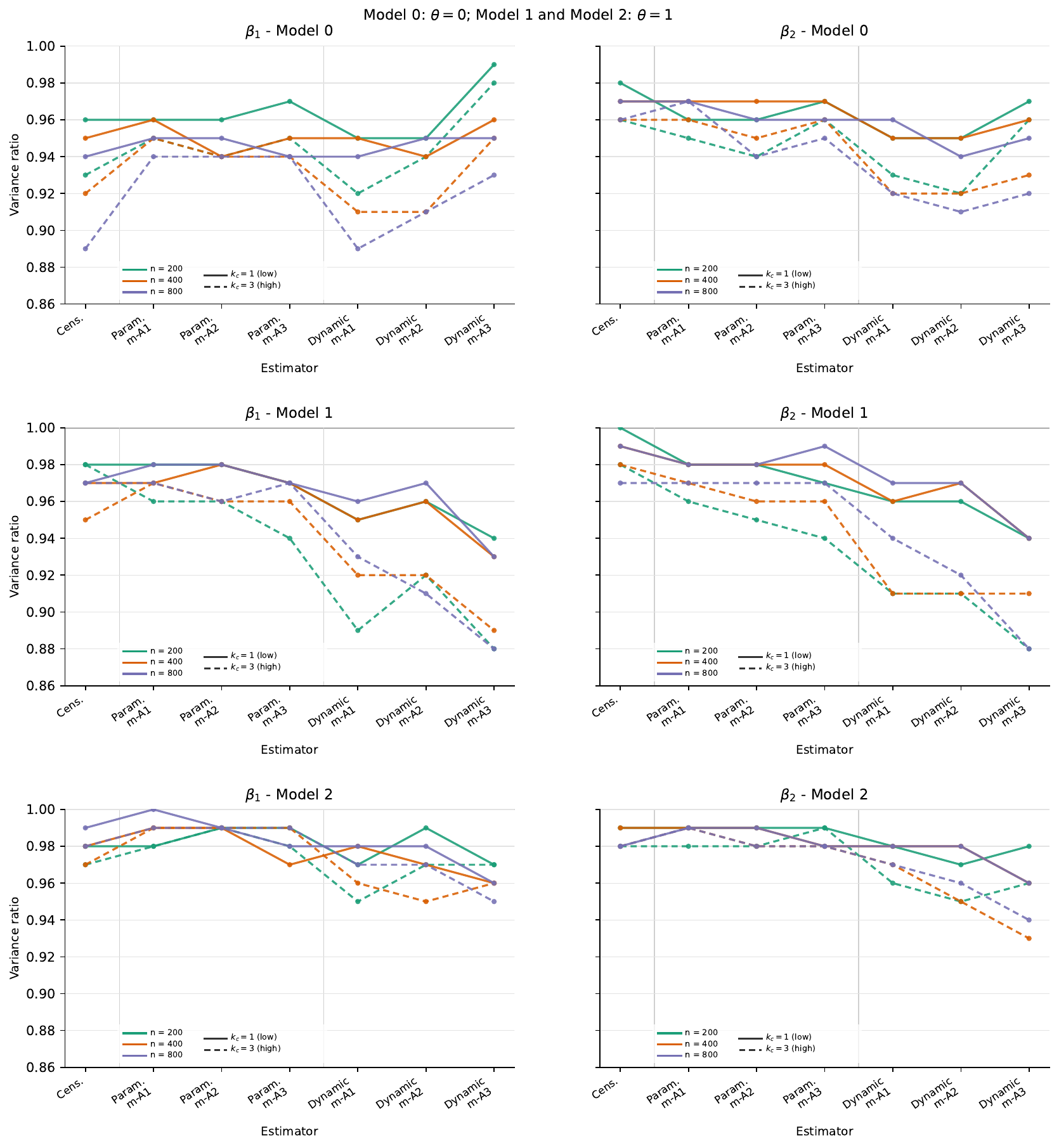}
    \caption{Variance ratio: Variance of the estimates from the adjusted estimators relative to the variance from the simple Ghosh-Lin estimator (\textit{base}), for $\theta=1$ in Models 1, 2, ($\theta=0$ in Model 0), for different censoring levels $k_c=1, 3$ and sample sizes $n$.}
    \label{fig:table-glvar-theta-1}
\end{figure}

\begin{figure}
    \centering
     \includegraphics[width=1.05\linewidth]{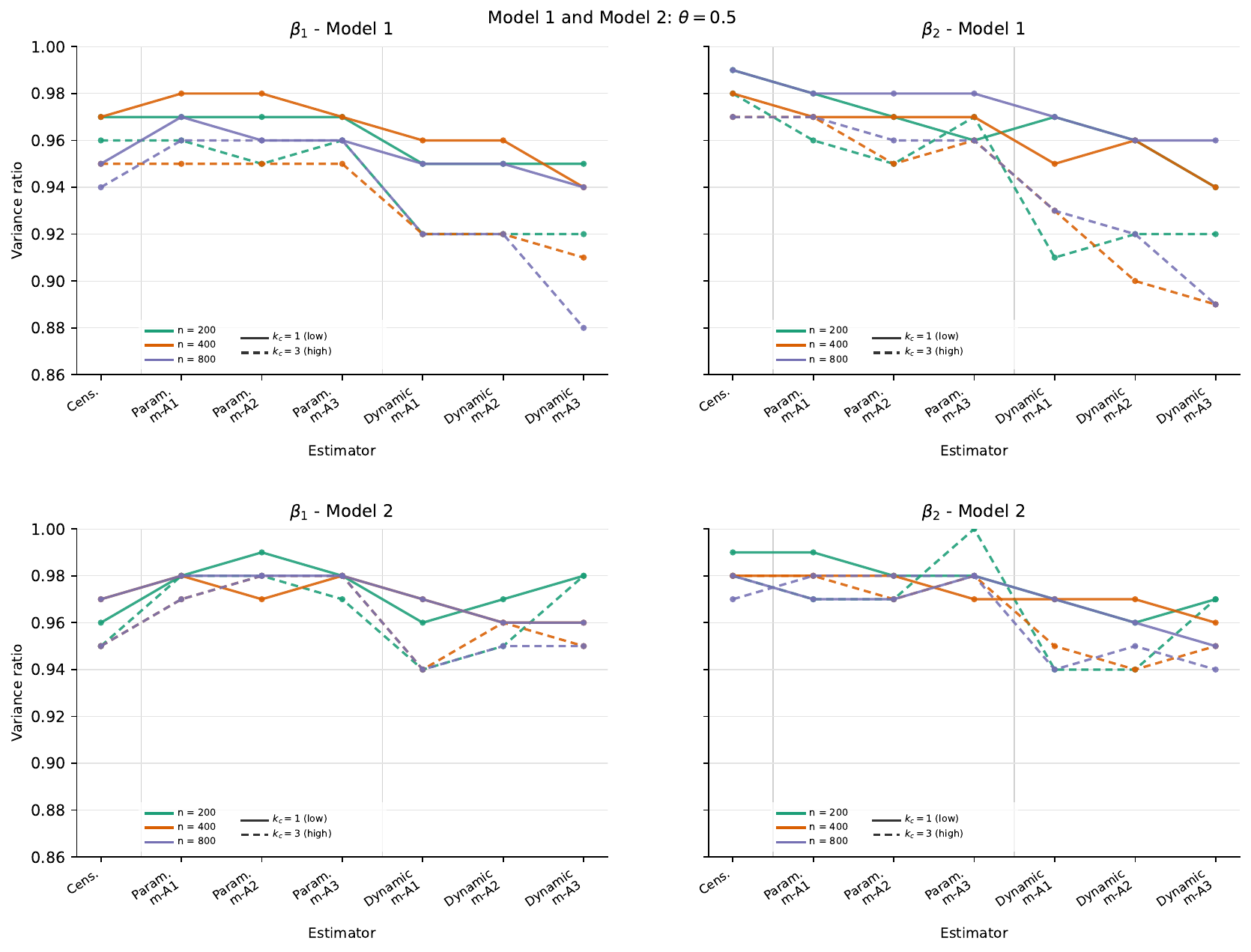}
    \caption{Variance ratio: Variance of the estimates from the adjusted estimators relative to the variance from the simple Ghosh-Lin estimator (\textit{base}), for $\theta=0.5$ in Models 1, 2, for different censoring levels $k_c=1, 3$ and sample sizes $n$.}
    \label{fig:table-glvar-theta-0.5}
\end{figure}

Figures \ref{fig:table-glvar-theta-1} and \ref{fig:table-glvar-theta-0.5} (equivalent to Web Table 1) show the variance of the estimates from the different adjusted estimators relative to the variance of estimates from the simple standard Ghosh-Lin estimator (\textit{base}). All variance ratios are below one, thus showing that the adjusted estimators provide an efficiency gain in all the considered cases.
In almost all settings, the efficiency gain increases with high censoring.
The dynamic-prediction augmentation provides the highest efficiency gain in all settings of Model 1 and Model 2, indicating that the variance reduction is achieved in any case, independently of the $m$-A prediction model, and it is always convenient to apply this estimation approach when there is heterogeneity in the data. In particular, under Model 1 and a stronger data heterogeneity ($\theta=1$), we observe the highest variance reductions for both regression coefficients, up to 12\% when the more complex dynamic model \(m\)-A3 is used. This conclusion is unchanged when $n$ decreases from 800 to $200$. 
Moreover, in all settings of Model 1,  the more complex dynamic model \(m\)-A3 is the best choice because it leads to a higher efficiency gain than the models \(m\)-A1 and \(m\)-A2, if we look at both coefficient estimators. However, under the weaker heterogeneity assumption ($\theta=0.5$) in Model 2, simulations show that the more parsimonious dynamic model \(m\)-A1 could be appropriate for small samples with a higher censoring level.

Under Model 0, that is, when there is no heterogeneity in the recurrent events and they are independent of the terminal events given the covariates, the IPCW stratification (\textit{cens}) appears to reduce the variance in particular for high censoring ($k_c=3$) and for the estimator $\hat\beta_1$ of the effect of the binary covariate $X_1$; the maximum efficiency gain is up to 11\% for the larger sample size $n=800$. 
Interestingly, under Model 0, the augmentation based on the simpler dynamic models \(m\)-A1 and \(m\)-A2 helps to reduce the variance of the coefficient estimator for the continuous variable $X_2$, while this task is more difficult to achieve with the IPCW based on the K-M stratified censoring estimator, which seems to lose information when using the categorised version $\tilde X_2$.

The parametric augmentation models \(m\)-A1, \(m\)-A2 and \(m\)-A3 do not show any advantage in variance reduction as compared to the dynamic versions, and in several settings they perform similarly to the estimation with simple IPCW stratification (\textit{cens}), and even worse in a few cases.   

Finally, in Table 3 the empirical coverage of our estimators is shown. Here the true value for $\beta$ (that solved the Ghosh-Lin estimating equation) was found by simulating a large
dataset repeatedly and fitting the model. We note that all estimators have a coverage that is close to the nominal level of 95 \% confidence intervals,
with only minor problems for the dynamic augmentation estimators that had
slightly too narrow bands for some settings and in particular for the smaller sample sizes and for the large censoring. 

One nice feature of the dynamic regression augmentation is that the implementation remains computationally manageable even with very large samples. It scales linearly in the implementation for all estimators, i.e., computing time grows approximately in proportion to the size of the data.  When, for example, a sample of 500000 subjects was generated, 
 this led to a data set with 1200000 records that was fitted
in 40 seconds for the dynamic regression augmented estimator with the more complex model m-3, although the current implementation is not particularly optimised.

\begin{table}[ht]
\centering
\begingroup
\renewcommand{\arraystretch}{0.80}
\scalebox{0.85}{
\begin{tabular}{cccc rr@{\hskip 0.3in} rrr r rrr}
\hline\noalign{\vskip\belowrulesep}
\multicolumn{1}{c}{} & \multicolumn{1}{c}{} & \multicolumn{1}{c}{} & \multicolumn{1}{c}{} & \textit{base} & \textit{cens} &  \multicolumn{3}{c}{\textit{Regression}-f(t,$\theta$)} & & \multicolumn{3}{c}{\textit{Dynamic Regression}} \\ 
\cmidrule(lr){7-9} \cmidrule(lr){11-13} 
 $k_c$ & Model & $\theta$ & Parameter  & \multicolumn{1}{c}{} & \multicolumn{1}{c}{} &  \(m\)-A1 & \(m\)-A2 & \(m\)-A3 & &  \(m\)-A1 & \(m\)-A2 & \(m\)-A3  \\ 
  \hline\noalign{\vskip\belowrulesep}
\multirow{10}{*}{1} & \multirow{2}{*}{0} & \multirow{2}{*}{0.0} & $\beta_1$  &  0.03 & 0.03 & 0.03 & 0.03 & 0.02  & & 0.03 & 0.02 & 0.02 \\ 
 & & & $\beta_2$ & -0.02 & -0.03 & -0.02 & -0.03 & -0.03 &  & -0.03 & -0.03 & -0.03 \\ \cmidrule(lr){2-13}
 & \multirow{4}{*}{1} & \multirow{2}{*}{0.5} & $\beta_1$ & 0.02 & 0.03 & 0.02 & 0.02 & 0.03 &  & 0.02 & 0.02 & 0.02 \\ 
 & & & $\beta_2$ & -0.02 & -0.02 & -0.02 & -0.03 & -0.03 &  & -0.02 & -0.03 & -0.03 \\ \cmidrule(lr){3-13}
 & & \multirow{2}{*}{1.0} & $\beta_1$ & 0.02 & 0.03 & 0.02 & 0.04 & 0.02 &  & 0.02 & 0.03 & 0.02 \\ 
 & & & $\beta_2$ & -0.02 & -0.02 & -0.02 & -0.03 & -0.03 &  & -0.02 & -0.03 & -0.03 \\ \cmidrule(lr){2-13}
 & \multirow{4}{*}{2} & \multirow{2}{*}{0.5} & $\beta_1$ & 0.05 & 0.05 & 0.05 & 0.05 & 0.05 &  & 0.05 & 0.04 & 0.05 \\ 
 & & & $\beta_2$ & -0.05 & -0.05 & -0.05 & -0.05 & -0.05 &  & -0.05 & -0.05 & -0.05 \\ \cmidrule(lr){3-13}
 & & \multirow{2}{*}{1.0} & $\beta_1$ & 0.06 & 0.06 & 0.06 & 0.06 & 0.06 &  & 0.06 & 0.06 & 0.06 \\ 
 & & & $\beta_2$ & -0.06 & -0.06 & -0.06 & -0.06 & -0.06 &  & -0.06 & -0.06 & -0.06 \\ 
   \hline\noalign{\vskip\belowrulesep}
\multirow{10}{*}{3} & \multirow{2}{*}{0} & \multirow{2}{*}{0.0} & $\beta_1$ & 0.09 & 0.09 & 0.09 & 0.09 & 0.09  & & 0.09 & 0.09 & 0.09 \\ 
 & & & $\beta_2$ & -0.09 & -0.09 & -0.09 & -0.09 & -0.09 &  & -0.09 & -0.09 & -0.10 \\ \cmidrule(lr){2-13}
 & \multirow{4}{*}{1} & \multirow{2}{*}{0.5} & $\beta_1$ & 0.09 & 0.09 & 0.09 & 0.09 & 0.10 &  & 0.09 & 0.09 & 0.10 \\ 
 & & & $\beta_2$ & -0.09 & -0.09 & -0.09 & -0.09 & -0.09 &  & -0.09 & -0.09 & -0.09 \\ \cmidrule(lr){3-13}
 & & \multirow{2}{*}{1.0} & $\beta_1$ & 0.08 & 0.09 & 0.09 & 0.10 & 0.09 &  & 0.09 & 0.09 & 0.09 \\ 
 & & & $\beta_2$ & -0.09 & -0.09 & -0.09 & -0.08 & -0.09 &  & -0.09 & -0.08 & -0.09 \\ \cmidrule(lr){2-13}
 & \multirow{4}{*}{2} & \multirow{2}{*}{0.5} & $\beta_1$ & 0.10 & 0.10 & 0.10 & 0.10 & 0.10 &  & 0.10 & 0.10 & 0.10 \\ 
 & & & $\beta_2$ & -0.10 & -0.10 & -0.10 & -0.10 & -0.10  & & -0.10 & -0.10 & -0.10 \\ \cmidrule(lr){3-13}
 & & \multirow{2}{*}{1.0} & $\beta_1$ & 0.10 & 0.10 & 0.10 & 0.10 & 0.10 &  & 0.10 & 0.10 & 0.10 \\ 
 & & & $\beta_2$ & -0.10 & -0.10 & -0.10 & -0.10 & -0.10 &  & -0.10 & -0.10 & -0.11 \\ 
   \hline
\end{tabular}
}
\endgroup
\caption{Mean of estimates ($n=200$). Data are simulated from Model 0, 1, 2, with different levels of dependence ($\theta$: variance of the frailty $Z$) and censoring rate ($k_c=1$: low, $k_c=3$: high).} 
\label{glmean}
\end{table}

\begin{table}[ht]
\centering
\begingroup
\renewcommand{\arraystretch}{0.80}
\scalebox{0.75}{
\begin{tabular}{ccccc rr@{\hskip 0.3in} rrr r rrr}
\hline\noalign{\vskip\belowrulesep}
\multicolumn{1}{c}{} & \multicolumn{1}{c}{} & \multicolumn{1}{c}{} &
\multicolumn{1}{c}{} & \multicolumn{1}{c}{} & \textit{base} & \textit{cens} &
\multicolumn{3}{c}{\textit{Regression}-$f(t,\theta)$} & &
\multicolumn{3}{c}{\textit{Dynamic Regression}} \\
\cmidrule(lr){8-10} \cmidrule(lr){12-14}
Model & $n$ & $k_c$  & $\theta$  & Parameter & \multicolumn{1}{c}{} &
\multicolumn{1}{c}{} & \(m\)-A1 & \(m\)-A2 & \(m\)-A3 & & \(m\)-A1 & \(m\)-A2 & \(m\)-A3 \\
\hline\noalign{\vskip\belowrulesep}
\multirow{12}{*}{0} & \multirow{4}{*}{200} & \multirow{2}{*}{1} & \multirow{2}{*}{0.0} & $\beta_1$ & 0.94 & 0.94 & 0.94 & 0.94 & 0.94 & & 0.94 & 0.94 & 0.93 \\ 
   &  &  &  & $\beta_2$  & 0.93 & 0.94 & 0.93 & 0.94 & 0.93 & & 0.93 & 0.94 & 0.93 \\ 
\cmidrule(lr){3-14}
  &   & \multirow{2}{*}{3}  & \multirow{2}{*}{0.0} & $\beta_1$ & 0.94 & 0.94 & 0.94 & 0.94 & 0.94 & & 0.94 & 0.94 & 0.93 \\ 
  &  &  &  & $\beta_2$  & 0.93 & 0.93 & 0.93 & 0.94 & 0.93 & & 0.93 & 0.94 & 0.92 \\ 
\cmidrule(lr){2-14}
  & \multirow{4}{*}{400}  & \multirow{2}{*}{1} & \multirow{2}{*}{0.0} & $\beta_1$  & 0.95 & 0.94 & 0.94 & 0.95 & 0.95 & & 0.94 & 0.94 & 0.95 \\ 
   &   &  &  & $\beta_2$  & 0.94 & 0.94 & 0.94 & 0.94 & 0.94 & & 0.94 & 0.94 & 0.94 \\ 
\cmidrule(lr){3-14}
   &  & \multirow{2}{*}{3}  & \multirow{2}{*}{0.0} & $\beta_1$  & 0.95 & 0.95 & 0.95 & 0.95 & 0.94 & & 0.94 & 0.95 & 0.93 \\ 
  &  &  &  & $\beta_2$  & 0.94 & 0.94 & 0.94 & 0.94 & 0.94 & & 0.94 & 0.94 & 0.93 \\ 
\cmidrule(lr){2-14}
  & \multirow{4}{*}{800}  & \multirow{2}{*}{1} & \multirow{2}{*}{0.0} & $\beta_1$  & 0.95 & 0.95 & 0.95 & 0.94 & 0.95 & & 0.95 & 0.94 & 0.95 \\ 
    &   &  &  & $\beta_2$   & 0.95 & 0.95 & 0.95 & 0.94 & 0.94 & & 0.95 & 0.94 & 0.94 \\ 
\cmidrule(lr){3-14}
  &  & \multirow{2}{*}{3}  & \multirow{2}{*}{0.0} & $\beta_1$ & 0.94 & 0.95 & 0.95 & 0.95 & 0.95 & & 0.95 & 0.95 & 0.95 \\ 
   &  &  &  & $\beta_2$ & 0.94 & 0.94 & 0.94 & 0.95 & 0.95 & & 0.94 & 0.95 & 0.95 \\ 
\cmidrule(lr){1-14}
\multirow{24}{*}{1} & \multirow{8}{*}{200} & \multirow{4}{*}{1} & \multirow{2}{*}{0.5} & $\beta_1$  & 0.94 & 0.94 & 0.94 & 0.94 & 0.94 & & 0.94 & 0.93 & 0.94 \\ 
   &  &  &  & $\beta_2$  & 0.93 & 0.93 & 0.93 & 0.93 & 0.94 & & 0.93 & 0.93 & 0.94 \\ 
  &   &   & \multirow{2}{*}{1} & $\beta_1$  & 0.93 & 0.94 & 0.93 & 0.93 & 0.93 & & 0.93 & 0.93 & 0.93 \\ 
   &  &  &  & $\beta_2$  & 0.92 & 0.92 & 0.92 & 0.93 & 0.93 & & 0.92 & 0.93 & 0.92 \\ 
\cmidrule(lr){3-14}
  &  & \multirow{4}{*}{3} & \multirow{2}{*}{0.5} & $\beta_1$ & 0.94 & 0.94 & 0.94 & 0.94 & 0.93 & & 0.93 & 0.93 & 0.92 \\ 
   &  &  &  & $\beta_2$ & 0.92 & 0.92 & 0.93 & 0.93 & 0.93 & & 0.92 & 0.93 & 0.91 \\ 
  &   &   & \multirow{2}{*}{1} & $\beta_1$  & 0.93 & 0.93 & 0.93 & 0.93 & 0.93 & & 0.93 & 0.93 & 0.92 \\ 
   &  &  &  & $\beta_2$ & 0.92 & 0.92 & 0.92 & 0.93 & 0.92 & & 0.92 & 0.93 & 0.91 \\ 
\cmidrule(lr){2-14}
  &  \multirow{8}{*}{400} & \multirow{4}{*}{1} & \multirow{2}{*}{0.5} & $\beta_1$  & 0.95 & 0.94 & 0.94 & 0.94 & 0.94 & & 0.94 & 0.94 & 0.94 \\ 
  &  &  &  & $\beta_2$  & 0.94 & 0.95 & 0.95 & 0.94 & 0.94 & & 0.95 & 0.94 & 0.94 \\ 
  &   &   & \multirow{2}{*}{1} & $\beta_1$ & 0.94 & 0.95 & 0.94 & 0.94 & 0.94 & & 0.95 & 0.94 & 0.94 \\ 
   &  &  &  & $\beta_2$  & 0.93 & 0.93 & 0.93 & 0.93 & 0.94 & & 0.93 & 0.94 & 0.94 \\ 
\cmidrule(lr){3-14}
  &  & \multirow{4}{*}{3} & \multirow{2}{*}{0.5} & $\beta_1$ & 0.94 & 0.94 & 0.94 & 0.95 & 0.94 & & 0.94 & 0.95 & 0.93 \\ 
    &  &  &  & $\beta_2$ & 0.94 & 0.94 & 0.94 & 0.94 & 0.94 & & 0.94 & 0.94 & 0.93 \\ 
  &   &   & \multirow{2}{*}{1} & $\beta_1$  & 0.93 & 0.93 & 0.93 & 0.94 & 0.93 & & 0.93 & 0.94 & 0.93 \\ 
   &  &  &  & $\beta_2$  & 0.93 & 0.93 & 0.93 & 0.94 & 0.94 & & 0.94 & 0.93 & 0.93 \\ 
\cmidrule(lr){2-14}
 &  \multirow{8}{*}{800} & \multirow{4}{*}{1} & \multirow{2}{*}{0.5} & $\beta_1$  & 0.94 & 0.94 & 0.94 & 0.95 & 0.95 & & 0.94 & 0.95 & 0.95 \\ 
   &  &  &  & $\beta_2$  & 0.94 & 0.94 & 0.94 & 0.94 & 0.94 & & 0.94 & 0.94 & 0.94 \\ 
  &   &   & \multirow{2}{*}{1} & $\beta_1$  & 0.95 & 0.95 & 0.94 & 0.94 & 0.94 & & 0.94 & 0.95 & 0.94 \\ 
   &  &  &  & $\beta_2$  & 0.94 & 0.94 & 0.94 & 0.94 & 0.94 & & 0.94 & 0.94 & 0.94 \\ 
\cmidrule(lr){3-14}
  &  & \multirow{4}{*}{3} & \multirow{2}{*}{0.5} & $\beta_1$ & 0.94 & 0.94 & 0.94 & 0.94 & 0.95 & & 0.94 & 0.94 & 0.95 \\ 
   &  &  &  & $\beta_2$  & 0.94 & 0.94 & 0.94 & 0.95 & 0.94 & & 0.94 & 0.95 & 0.94 \\ 
  &   &   &  \multirow{2}{*}{1} & $\beta_1$ & 0.94 & 0.94 & 0.95 & 0.94 & 0.94 & & 0.94 & 0.94 & 0.93 \\ 
  &  &  &  & $\beta_2$ & 0.94 & 0.94 & 0.94 & 0.94 & 0.94 & & 0.94 & 0.94 & 0.94 \\ 
\cmidrule(lr){1-14}
\multirow{24}{*}{2} & \multirow{8}{*}{200} & \multirow{4}{*}{1} & \multirow{2}{*}{0.5} & $\beta_1$   & 0.94 & 0.94 & 0.94 & 0.93 & 0.94 & & 0.94 & 0.93 & 0.94 \\ 
   &  &  &  & $\beta_2$  & 0.94 & 0.94 & 0.94 & 0.94 & 0.93 & & 0.94 & 0.94 & 0.93 \\ 
  &   &   & \multirow{2}{*}{1} & $\beta_1$   & 0.94 & 0.94 & 0.94 & 0.94 & 0.94 & & 0.94 & 0.94 & 0.94 \\ 
  &  &  &  & $\beta_2$  & 0.94 & 0.94 & 0.94 & 0.93 & 0.94 & & 0.94 & 0.93 & 0.94 \\ 
\cmidrule(lr){3-14}
  &   & \multirow{4}{*}{3} & \multirow{2}{*}{0.5} & $\beta_1$   & 0.95 & 0.95 & 0.95 & 0.94 & 0.94 & & 0.94 & 0.93 & 0.92 \\ 
   &  &  &  & $\beta_2$   & 0.93 & 0.93 & 0.93 & 0.93 & 0.93 & & 0.93 & 0.93 & 0.92 \\ 
  &   &   & \multirow{2}{*}{1} & $\beta_1$   & 0.94 & 0.94 & 0.94 & 0.94 & 0.94 & & 0.94 & 0.94 & 0.93 \\ 
  &  &  &  & $\beta_2$   & 0.93 & 0.93 & 0.93 & 0.93 & 0.93 & & 0.93 & 0.93 & 0.92 \\ 
\cmidrule(lr){2-14}
  & \multirow{8}{*}{400} & \multirow{4}{*}{1} & \multirow{2}{*}{0.5} & $\beta_1$ & 0.94 & 0.94 & 0.94 & 0.95 & 0.94 & & 0.94 & 0.95 & 0.95 \\ 
  &  &  &  & $\beta_2$  & 0.94 & 0.94 & 0.94 & 0.95 & 0.94 & & 0.94 & 0.95 & 0.94 \\ 
 &  &  & \multirow{2}{*}{1} & $\beta_1$  & 0.95 & 0.95 & 0.95 & 0.94 & 0.95 & & 0.95 & 0.94 & 0.94 \\ 
  &  &  &  & $\beta_2$ & 0.94 & 0.94 & 0.94 & 0.94 & 0.93 & & 0.94 & 0.94 & 0.93 \\ 
\cmidrule(lr){3-14}
 &  & \multirow{4}{*}{3} & \multirow{2}{*}{0.5} & $\beta_1$  & 0.95 & 0.95 & 0.94 & 0.94 & 0.94 & & 0.95 & 0.94 & 0.94 \\ 
 &  &  &  & $\beta_2$  & 0.94 & 0.94 & 0.94 & 0.94 & 0.94 & & 0.94 & 0.94 & 0.94 \\ 
  &   &   & \multirow{2}{*}{1} & $\beta_1$  & 0.95 & 0.95 & 0.95 & 0.94 & 0.95 & & 0.95 & 0.94 & 0.94 \\ 
  &  &  &  & $\beta_2$  & 0.94 & 0.94 & 0.94 & 0.93 & 0.94 & & 0.94 & 0.93 & 0.94 \\ 
\cmidrule(lr){2-14}
  & \multirow{8}{*}{800} & \multirow{4}{*}{1} & \multirow{2}{*}{0.5} & $\beta_1$ & 0.95 & 0.95 & 0.95 & 0.94 & 0.94 & & 0.95 & 0.95 & 0.94 \\ 
   &  &  &  & $\beta_2$  & 0.95 & 0.95 & 0.95 & 0.95 & 0.94 & & 0.95 & 0.95 & 0.94 \\ 
  &   &   & \multirow{2}{*}{1} & $\beta_1$ & 0.95 & 0.95 & 0.94 & 0.95 & 0.95 & & 0.94 & 0.95 & 0.95 \\ 
 &  &  &  & $\beta_2$  & 0.94 & 0.94 & 0.94 & 0.95 & 0.94 & &  0.94 & 0.95 & 0.94 \\ 
 \cmidrule(lr){3-14}
 &  & \multirow{4}{*}{3} & \multirow{2}{*}{0.5} & $\beta_1$   & 0.95 & 0.95 & 0.95 & 0.95 & 0.95 & & 0.95 & 0.95 & 0.95 \\ 
 &  &  &  & $\beta_2$  & 0.95 & 0.95 & 0.95 & 0.94 & 0.95 & & 0.94 & 0.94 & 0.94 \\ 
  &   &   & \multirow{2}{*}{1} & $\beta_1$ & 0.95 & 0.94 & 0.95 & 0.95 & 0.95 & & 0.94 & 0.95 & 0.95 \\ 
&  &  &  & $\beta_2$  & 0.94 & 0.94 & 0.95 & 0.95 & 0.95 & & 0.94 & 0.95 & 0.95 \\ 
   \hline
\end{tabular}
}
\endgroup
\caption{ Empirical coverage of 95\% confidence intervals for different censoring levels $k_c$ and sample sizes $n$. Data are simulated from Models 0, 1, 2 with frailty variance $\theta$.} 
\label{glcov}
\end{table}

\subsection{Misspecified Censoring Distribution}

To show the bias present when the censoring model is misspecified, we simulated data in a slightly different way to guarantee that the true parameters of the Ghosh-Lin model were known. 
We considered two i.i.d.  Bernoulli covariates $X=(X_1, X_2)$ with $P(X_1=1)=P(X_2=1)=0.5$.
Survival times were simulated as in the main Section \ref{sim}, via a Cox model with the same coefficients $\beta_d=(0.3,-0.3)$ and baseline as before. On the other hand, recurrent events were simulated assuming that the rate for survivors is given as 
$\lambda_{0}(t) \exp(X^T \beta) / S(t,X)$ with $\beta=(0.3,-0.3)$,
such that, by construction, $E(N(t \wedge D) | X) = \Lambda_0(t) \exp(X^T \beta)$, that is, a Ghosh-Lin model with the true parameters  $\beta$. We also considered extensions of this
model that included random effects, but here we only report results from the setting with independence. 
The censoring times were generated from a Cox regression model with hazard $(3/5000) \cdot \exp(X^T \beta_c)$ (high censoring level $k_c=3$) under three settings:  
$\beta_c=(0,0)$ (independent),  $\beta_c=(0.5,0)$ ($X_1$-Censoring), $\beta_c=(0,0.5)$ ($X_2$-Censoring).

For this setting, Table \ref{biasGL} reports the mean parameter estimates across 5000 simulation replications with a sample size of $n=1600$. 
As a reference, under covariate--independent censoring situation, all estimators are approximately unbiased for the effects of both covariates. 
When censoring depends on $X_1$, the standard estimator (`\textit{base}') is severely biased in estimating the effect of $X_1$, and so is the parametric-regression augmented estimator.  
As expected, the IPCW censoring-adjusted estimator (`\textit{cens}') is essentially unbiased in this setting.  
Surprisingly, the dynamic-regression augmented estimators show only modest bias. 
A similar pattern emerges when censoring depends on $X_2$: the standard estimator and the parametric-regression augmented estimator exhibits substantial bias in estimating the effect of $X_2$, whereas the IPCW censoring-adjusted estimator remained approximately unbiased. Again, the dynamic-regression augmented estimators exhibited only little bias.

\begin{table}[ht]
\centering
\begingroup
\renewcommand{\arraystretch}{0.80}
\scalebox{0.85}{
\begin{tabular}{ccrr@{\hskip 0.3in} rrr r rrr}
\hline\noalign{\vskip\belowrulesep}
 \multicolumn{1}{c}{}  & \multicolumn{1}{c}{}  & \textit{base} & \textit{cens} & \multicolumn{3}{c}{\textit{Regression-f(t,$\theta$)}} &  &  \multicolumn{3}{c}{\textit{Dynamic Regression}} \\  
 \cmidrule(lr){5-7} \cmidrule(lr){9-11}
 Assumption &   Parameter & \multicolumn{1}{c}{} &   \multicolumn{1}{c}{} &  \(m\)-A1 & \(m\)-A2 & \(m\)-A3 &  & \(m\)-A1 & \(m\)-A2 & \(m\)-A3  \\ 
\hline\noalign{\vskip\belowrulesep}
\multirow{2}{*}{Independent} & $\beta_1$ & 0.30 & 0.30 & 0.30 & 0.30 & 0.30 &  & 0.30 & 0.30 & 0.29 \\ 
            & $\beta_2$ & -0.30 & -0.30 & -0.30 & -0.31 & -0.30 &  & -0.30 & -0.30 & -0.29 \\ 
\hline\noalign{\vskip\belowrulesep}
\multirow{2}{*}{$X_1$-Censoring}  & $\beta_1$ & 0.07 & 0.29 & 0.19 & 0.20 & 0.20 &  & 0.26 & 0.25 & 0.25 \\ 
       & $\beta_2$ & -0.31 & -0.32 & -0.31 & -0.31 & -0.30 &  & -0.31 & -0.30 & -0.28 \\ 
\hline\noalign{\vskip\belowrulesep}
\multirow{2}{*}{$X_2$-Censoring}   & $\beta_1$ & 0.31 & 0.30 & 0.30 & 0.30 & 0.30 &  & 0.29 & 0.29 & 0.28 \\ 
       & $\beta_2$ & -0.52 & -0.32 & -0.42 & -0.43 & -0.42 &  & -0.34 & -0.35 & -0.35 \\ 
   \hline
\end{tabular}
}
\endgroup
\caption{Mean of estimates in the dependent censoring case. Data were simulated under independent censoring, censoring dependent on $X_1$, and censoring dependent on $X_2$.} 
\label{biasGL}
\end{table}

\subsection{A randomized clinical study with auxiliary covariates}

In a complementary simpler simulation study, we focus exclusively on estimation of the effect of the binary covariate $X_1$ on the recurrent event rate, a typical setting where, for example, a treatment of interest is investigated in a randomized clinical study (RCT). 

Data were generated from the two-stage recurrent-events random effects model of \citet{Scheike2026}, in
which the recurrent event process and the terminal event share a common random
frailty, inducing dependence both among successive recurrent events and between the
recurrent events and the terminal event; this corresponds to Model~2 above. The
frailty distribution is Gamma with mean $1$ and variance $\theta$. 
This
construction leads to a terminal event rate on the proportional hazards form
$\lambda_d(t)=\alpha_d(t)\exp(X_1\beta_{d1} + X_2 \beta_{d2})$, while the recurrent event 
process follows
a Ghosh-Lin proportional means model for $E(N^*(t)|X_1,X_2)$ with cumulative mean function
$\Lambda_0(t)\exp(X_1\beta_{1} + X_2 \beta_{2} )$, with 
$\beta_{1}=\beta_{d1}=0.3$ and
$\beta_{2}=\beta_{d2}=-0.3$.
The covariate $X_1$ is
Binomial with $P(X_1=1)=0.5$ and independent of $X_2$, which has a standard normal distribution. Both covariates are generated independently of the censoring mechanism,
mirroring a randomized controlled trial in which $X_1$ plays the role of a randomized
treatment indicator, while the recurrent event count $N(t-)$ and $X_2$, and possibly also the frailty $Z$, represent auxiliary
information that can be exploited by the augmented estimators purely to gain
efficiency in estimating the effect of $X_1$, without introducing bias.  The baseline
intensity/mean functions are taken from the application (see next section) and set as nonparametric estimates of the recurrent CRBSI
event rate and the terminal event rate obtained from a real central-catheter cohort,
scaled by a factor of $0.5$, so that the simulated event histories mimic a
lower-intensity version of this specific application $X_1$ (see Figure \ref{fig:cumrate_sim}). The censoring time is exponentially distributed with hazard $\lambda_c=3/5000$.

Since $X_2$ is generated independently of $X_1$ and
censoring is independent of both $X_1$ and $X_2$, the proportional means model for $E(N^*(t)|X_1)$ given  only $X_1$  is also on Ghosh-Lin form mixing out over $X_2$, so also the true effect of $X_1$  in this model 
is $\beta_{1}=0.3$ (while the baseline cumulative mean depends on $X_2$). 

For each simulated dataset, we fit five Ghosh-Lin models for the mean number of recurrent events, in each case with covariate $X_1$ only
(mimicking an RCT setting in which $X_1$ is the randomized treatment and $X_2$
is a baseline auxiliary covariate not adjusted for in the primary model), and
extract only the estimated coefficient of $X_1$ from each: (i) the standard unaugmented Ghosh-Lin model, with 
unstratified IPCW censoring weights (\emph{base}); (ii) the same model with censoring weights
stratified on $(X_1,\tilde X_2)$, where $\tilde X_2 = \mathbb{I}\{X_2>0\}$ is
$X_2$ dichotomized at zero (\emph{cens-$X_1\tilde X_2$}); (iii) an augmented
estimator using the static regression augmentation with prediction set
$e(t)=(X_1,X_2)$ (\emph{aug-$X_1X_2$}); (iv) an augmented estimator with
prediction set $e(t)=(X_1,X_2,N(t^-))$ (\emph{aug-$X_1X_2N(t^-)$}); and (v) an
augmented estimator with prediction set $e(t)=(X_1,X_2,Z,N(t^-))$
(\emph{aug-$X_1X_2ZN(t^-)$}), where $Z$ denotes the (in practice unobserved)
frailty driving the recurrent event process. Only the static regression
augmentation is used for (iii)--(v), and, as before, the augmentation term is
computed once and the model is then refit using it as a fixed adjustment.
Estimator (v), which augments on the true frailty in addition to $X_1$, $X_2$,
and $N(t^-)$, is included solely as an oracle benchmark representing the best
achievable efficiency under full information, rather than as a practically implementable estimator. 
 
The recurrent event process
is by construction of Ghosh-Lin form in $X_1$ with true coefficient
$\beta_1=0.3$, so all five estimators target the same known value $\beta_1$. 
The
comparison across them therefore isolates the efficiency gain achieved by
exploiting auxiliary information on $X_2$ (and, in the oracle case, the
frailty $Z$, together with $N(t^-)$) through augmentation and through
stratified censoring weights, without altering the primary model
specification---in a setting directly analogous to efficiency gains from
baseline covariate adjustment in a randomized trial. 
For each of the five
estimators, in Table \ref{tab:auxcov} we report the empirical mean and standard deviation of
$\hat\beta_1$ over replications, the mean model-based standard error, 
and the empirical coverage of the 95\% confidence interval for the true $\beta_1$, based on 2000 independent replications for a chosen sample
size $n$.

\begin{table}[ht]
\centering
\scalebox{0.95}{
\begin{tabular}{ccl rrrrr}
\hline\noalign{\vskip\belowrulesep}
$\theta$  & $n$  & Estimator &  Mean & Bias & EmpSD & MeanSE & Coverage \\ 
\hline\noalign{\vskip\belowrulesep}
\multirow{10}{*}{1} & \multirow{5}{*}{400}  & \textit{Base} 
  & 0.309 & 0.009 & 0.199 & 0.189 & 0.930 \\ 
   &  & \emph{cens-$X_1\tilde X_2$}  & 0.310 & 0.010 & 0.193 & 0.183 & 0.932 \\ 
   &  & \emph{aug-$X_1X_2$}  & 0.307 & 0.007 & 0.194 & 0.183 & 0.926 \\ 
    &  &  \emph{aug-$X_1X_2N(t^-)$} & 0.301 & 0.001 & 0.189 & 0.180 & 0.929 \\ 
    &  &  \emph{aug-$X_1X_2ZN(t^-)$} & 0.302 & 0.002 & 0.186 & 0.178 & 0.928 \\ 
    \cmidrule(lr){2-8}
  & \multirow{5}{*}{800}  & \textit{Base} 
  & 0.298 & -0.002 & 0.142 & 0.137 & 0.942 \\ 
   &  & \emph{cens-$X_1\tilde X_2$}  & 0.298 & -0.002 & 0.137 & 0.132 & 0.947 \\ 
   &  & \emph{aug-$X_1X_2$}   & 0.296 & -0.004 & 0.139 & 0.133 & 0.944 \\ 
    &  &  \emph{aug-$X_1X_2N(t^-)$}  & 0.291 & -0.009 & 0.136 & 0.132 & 0.942 \\ 
   &  &  \emph{aug-$X_1X_2ZN(t^-)$}  & 0.291 & -0.009 & 0.136 & 0.130 & 0.943 \\ 
   \cmidrule(lr){2-8}
    & \multirow{5}{*}{1600}  & \textit{Base} 
   & 0.303 & 0.003 & 0.099 & 0.098 & 0.950 \\ 
  &  & \emph{cens-$X_1\tilde X_2$}   & 0.303 & 0.003 & 0.096 & 0.095 & 0.949 \\ 
  &  & \emph{aug-$X_1X_2$}   & 0.302 & 0.002 & 0.097 & 0.096 & 0.949 \\ 
    &  &  \emph{aug-$X_1X_2N(t^-)$}  & 0.299 & -0.001 & 0.095 & 0.095 & 0.952 \\ 
  &  &  \emph{aug-$X_1X_2ZN(t^-)$}   & 0.298 & -0.002 & 0.094 & 0.094 & 0.949 \\ 
  \midrule
\multirow{10}{*}{2} & \multirow{5}{*}{400}  & \textit{Base}  & 0.304 & 0.004 & 0.223 & 0.211 & 0.943 \\ 
 &  & \emph{cens-$X_1\tilde X_2$}  & 0.303 & 0.003 & 0.217 & 0.206 & 0.946 \\ 
  &  & \emph{aug-$X_1X_2$}  & 0.301 & 0.001 & 0.217 & 0.205 & 0.938 \\ 
    &  &  \emph{aug-$X_1X_2N(t^-)$}  & 0.296 & -0.004 & 0.211 & 0.202 & 0.937 \\ 
  &  &  \emph{aug-$X_1X_2ZN(t^-)$}  & 0.299 & -0.001 & 0.208 & 0.199 & 0.936 \\ 
 \cmidrule(lr){2-8}
& \multirow{5}{*}{800}  & \textit{Base}   & 0.302 & 0.002 & 0.160  & 0.154 & 0.940 \\ 
 &  & \emph{cens-$X_1\tilde X_2$}  & 0.303 & 0.003 & 0.156 & 0.150 & 0.943 \\ 
  &  & \emph{aug-$X_1X_2$}   & 0.302 & 0.002 & 0.157 & 0.150 & 0.937 \\ 
   &  &  \emph{aug-$X_1X_2N(t^-)$}  & 0.298 & -0.002 & 0.154 & 0.148 & 0.937 \\ 
  &  &  \emph{aug-$X_1X_2ZN(t^-)$}   & 0.299 & -0.001 & 0.153 & 0.146 & 0.939 \\  
 \cmidrule(lr){2-8}
 & \multirow{5}{*}{1600}  & \textit{Base}  & 0.299 & -0.001 & 0.114 & 0.111 & 0.941 \\ 
  &  & \emph{cens-$X_1\tilde X_2$}  & 0.299 & -0.001 & 0.112 & 0.108 & 0.943 \\ 
   &  & \emph{aug-$X_1X_2$}   & 0.299 & -0.001 & 0.112 & 0.109 & 0.941 \\ 
    &  &  \emph{aug-$X_1X_2N(t^-)$}  & 0.295 & -0.005 & 0.110 & 0.108 & 0.938 \\ 
  &  &  \emph{aug-$X_1X_2ZN(t^-)$}   & 0.296 & -0.004 & 0.108 & 0.107 & 0.938 \\ 
   \bottomrule
\end{tabular}
}
\caption{{\small Simulation results across different sample sizes and frailty variance $\theta$. Data are simulated under Model 2, with true $\beta_1 = 0.3$ equal in all scenarios. EmpSD: empirical standard deviation of $\hat\beta_1$; MeanSE: mean model-based standard error; Coverage: empirical coverage of 95\% confidence interval.}}
\label{tab:auxcov}
\end{table}

Across all combinations of frailty variance ($\theta=1,2$) and sample
size ($n =400, 800, 1600$), the five estimators are essentially
unbiased for $\beta_1$, with any small residual bias attenuating as $n$
increases. The efficiency ordering is consistent across every scenario: the
unaugmented \emph{base} estimator has the largest empirical standard
deviation (EmpSD); \emph{cens-$X_1\tilde X_2$} and \emph{aug-$X_1X_2$}
perform very similarly to one another, confirming that stratifying the
censoring weights on $(X_1,\tilde X_2)$ and augmenting on $(X_1,X_2)$ extract
essentially the same efficiency gain from the auxiliary covariate $X_2$.
Adding $N(t^-)$ to the augmentation set yields a further, consistent
reduction in EmpSD, and the oracle estimator \emph{aug-$X_1X_2ZN(t^-)$},
which additionally exploits the true frailty, achieves the smallest EmpSD in
every scenario, establishing the efficiency bound attainable under full
information. As expected, all efficiency gaps narrow as $n$ increases, and
are more pronounced at $\theta=2$ than at $\theta=1$, consistent with a
larger frailty variance leaving more residual heterogeneity in $N(t^-)$ and
$Z$ for the augmentation to exploit.

The model-based standard errors (MeanSE) track EmpSD closely throughout,
confirming that the sandwich-type variance estimator is well calibrated
regardless of augmentation scheme or sample size. Empirical coverage of the Wald confidence intervals is close to the nominal level of 95\%, with no
evidence of severe under-coverage. For $\theta=1$, coverage improves toward
95\% as $n$ increases, consistent with the asymptotic justification of the
variance estimators. For $\theta=2$, coverage is already close to the nominal level at
the smallest sample size and remains stable when the sample size increases, with a slight tendency toward mild under-coverage for the more heavily augmented
estimators; this discrepancy is modest and does not
offset the substantial efficiency gains achieved.

\subsection{Low action setting}

Additional simulations considered a low action setting, with recurrent event and death rates scaled by factors of $0.1$ and $0.5$, respectively. This led to $37$ \% survivors for the baseline group, and an average number of recurrent events of $0.4$ at $5000$ days. All other parameters were set as in the previous subsection. 

 Simulation results are reported in Web Tables 2, 3 and 4 in the online Supplementary Material.
 The mean coefficient estimates are similar across the estimation methods. However, due to the lower death rate, the parameters change less across the different settings, as compared to the previous simulation study. 
 The benefit of the augmentation procedure is lost in this setting, providing little or no variance reduction relative to the standard Ghosh--Lin estimator. A slight efficiency gain is observed only in some settings of Models 0 and 1.
 Mild variance inflation occurred in some settings, particularly under Model 2, where a large heterogeneity is assumed, and with the more complex prediction model \(m\)-A3. 
  Empirical coverage was generally close to the nominal $95\%$ level in larger samples, although some undercoverage occurred at $n=200$, especially with prediction model \(m\)-A3 and under heavier censoring.

 These findings suggest that, when recurrent events are sparse, we learn less about individual heterogeneity; many subjects have \(N(t-)=0\), and the recurrence events history provides less information to distinguish their future risk. Therefore, estimating the augmentation coefficients, particularly interactions and the \(\gamma(t)\) coefficients in \(m\)-A3, becomes more difficult.
 In conclusion, when the recurrent event and death rates are very low, 
limited predictive information available from the event histories may be insufficient to offset the variability introduced by estimating the augmentation prediction models.

\section{Application}

Patients with chronic intestinal failure receiving home parenteral support (HPS)
through a central venous catheter can experience several complications during
the often long-term treatment periods, see
\citet{Tribler2018,cortese-scheike-2022}.  We consider a cohort of 715
consecutive patients at the University Hospital of Copenhagen, where we analysed the number of catheter-related bloodstream infections (CRBSI).  Some
of the patients died due to severe intestinal failure or co-morbidities, while
other patients left the HPS program alive for different reasons during the
around 14 years of follow-up.  We here studied the mean number of CBRSI's in
the first 5000 days of the follow-up, where only about $20$ \% of the patients
survived.  There is considerable heterogeneity between the patients with some
having 0 infections and others up to 43 infections.

We then described the predictive effect of the diagnosis adjusted for age and
sex, using the standard Ghosh-Lin model (`Standard Ghosh-Lin'), a Ghosh-Lin model with stratified
Kaplan-Meier censoring weights based on diagnosis and sex (`Ghosh-Lin stratified KM'), 
and two different linearly augmented Ghosh-Lin models (`Ghosh-Lin Augmented-1/2'). In the `Augmented-1' model,
only the main effects of the recurrent events count and all covariates are considered, while the `Augmented-2' model contains all two-way interactions and the same covariates as in `Augmented-1'.

The results in Table \ref{catheter} show that Diagnosis was significantly associated with the number of infections. Compared with patients diagnosed with `Cancer', those with `Other cause of intestinal failure' or `Inflammatory bowel disease' had 2.37-fold and 2.16-fold increases in the expected number of bloodstream infections, with ratios of 2.37 and 2.16, respectively.
All methods give quite similar estimates for the regression coefficients, but as expected, we do
gain in efficiency by using the stratified censoring weights and further by the
linear regression augmentation that uses the covariates information and the recurrent events.
There was some indication that the censoring distribution was dependent on diagnosis, so therefore all
augmented models were fitted using censoring weights stratified on diagnosis. The gain from this adjustment was considerable.

\begin{table}[ht]
\centering
\begingroup\small
\begin{tabular}{l ccc c ccc }
\hline\noalign{\vskip\belowrulesep}
	& \multicolumn{3}{c}{Standard Ghosh-Lin} & & \multicolumn{3}{c}{Ghosh-Lin stratified KM} \\  
    \cmidrule(lr){2-8}
 Covariate   & Estimate & SE & P-value & & Estimate & SE & P-value \\ 
\midrule
Diagnosis:  &  &  &  &  &  &  &  \\
\quad \textit{Complication at surgery} & 0.249 & 0.176 & 0.158 & & 0.215 & 0.174 & 0.215 \\ 
\quad \textit{Mesenteric vascular disease} & 0.181 & 0.191 & 0.345 & & 0.101 & 0.194 & 0.604 \\ 
\quad  \textit{Other cause of intestinal failure} & 0.862 & 0.342 & 0.012 & & 0.772 & 0.334 & 0.021 \\ 
\quad \textit{Inflammatory bowel disease} & 0.769 & 0.184 & 0.000 & & 0.682 & 0.181 & 0.000 \\ 
  Sex & -0.075 & 0.153 & 0.624 & & -0.082 & 0.153 & 0.595 \\ 
  Age & -0.007 & 0.005 & 0.210 & & -0.007 & 0.005 & 0.221 \\ 
\midrule
   &  \multicolumn{3}{c}{Ghosh-Lin Augmented-1} &  & \multicolumn{3}{c}{Ghosh-Lin Augmented-2} \\  
      \cmidrule(lr){2-8}
 Covariate   & Estimate & SE & P-value &  & Estimate & SE & P-value \\ 
\midrule
Diagnosis:  &  &  &  &  &  &  &  \\
\quad \textit{Complication at surgery} & 0.229 & 0.172 & 0.182 & & 0.249 & 0.167 & 0.137 \\ 
\quad \textit{Mesenteric vascular disease}  & 0.143 & 0.191 & 0.453 & & 0.162 & 0.186 & 0.384 \\ 
\quad \textit{Other cause of intestinal failure} & 0.781 & 0.317 & 0.014 & & 0.519 & 0.340 & 0.126 \\ 
\quad \textit{Inflammatory bowel disease} & 0.741 & 0.177 & 0.000 & & 0.736 & 0.174 & 0.000 \\ 
  Sex & -0.057 & 0.147 & 0.697 & & -0.016 & 0.142 & 0.912 \\ 
  Age & -0.004 & 0.005 & 0.470 & & -0.005 & 0.005 & 0.343 \\ 
   \hline
  \end{tabular}
\endgroup
\caption{Results on regression coefficients of different Ghosh-Lin models to study the number of catheter-related bloodstream infections (CRBSI) in patients with chronic intestinal failure. The reference category for Diagnosis is `Cancer'.} 
\label{catheter}
\end{table}

\section{Discussion} 

The paper develops a dynamic censoring-augmented estimator for the Ghosh–Lin proportional means model for recurrent events in presence of a terminal event to improve estimation of the marginal regression effects and obtain a variance reduction. 

The proposed framework has two main advantages. First, when the conditional mean prediction in the augmentation term is known, the censoring augmentation yields the fully efficient estimator with the smallest asymptotic variance in the corresponding class of regular estimating equations. Second, the dynamic-regression augmented estimator provides a practical approximation by projecting this prediction onto a chosen set of predictable regressors that include baseline covariates and the observed recurrent event history $N(t-)$, with either nonparametric time-varying coefficients or a simpler parametric time structure. Auxiliary covariates can also contribute to the augmentation without being included in the primary marginal mean model. 
The great advantage of the dynamic proposed estimators is that information collected before censoring can improve precision of estimates without requiring any additional assumption or a full joint model for recurrent and terminal events (e.g., as in frailty models).
The low computational cost for implementation also makes the approach feasible for large recurrent event data sets.

Moreover, under the assumed conditional independent censoring, a 
correct specification of the censoring model ensures consistency even when the dynamic-regression augmentation model used to approximate the conditional mean prediction is misspecified. With the augmentation coefficients estimated as described in the paper, we always get an efficiency gain, and the asymptotic variance of the proposed estimator is always lower than that of the standard Ghosh-Lin estimator. Of course, the size of the reduction gain depends on how well the selected predictors in the augmentation model explain the future contributions of the censored observations to the estimating equation. Then, 
a better working augmentation model can yield a larger variance reduction.  This robustness property is particularly useful in recurrent event settings where the joint dependence of the recurrent event history and the terminal event is often difficult to model reliably.
Finally, the variance reduction also holds when the Ghosh-Lin model is misspecified. In this case, the standard and augmented estimators target the same limiting working-model parameter. A further robustness result concerns randomized clinical trials: under the censoring conditions established for this setting, the treatment coefficient still converges to zero under the null hypothesis, even when the marginal mean model is misspecified. Combined with robust standard errors, this result allows valid testing of the null treatment effect.

The role of the censoring model must nevertheless be distinguished from that of the working augmentation model. General consistency still requires correct adjustment for the conditional censoring mechanism. The simulations with covariate-dependent censoring illustrate this point: when the censoring model omitted the relevant covariate, and thus it is misspecified, the standard estimator was substantially biased, whereas the estimator using appropriately stratified IPCW was approximately unbiased. Dynamic augmentation considerably attenuated the bias in the examined scenarios, but it should not be interpreted as a general substitute for a well-specified censoring model. This empirical bias reduction should be distinguished from a general double robustness property for the fitted dynamic regression estimator. Correct specification of the censoring model remains important for the general consistency result.

The simulation results support these theoretical findings. The dynamic augmentation reduced the empirical variance relative to the standard Ghosh-Lin estimator in all considered scenarios. The largest gains, up to about 12\%, occurred with stronger heterogeneity and heavier censoring, when the observed event history carries more information about future events. Time-varying dynamic models were generally preferable to the more restricted parametric time structure. The additional randomized-trial simulations also showed that auxiliary baseline covariates, together with the history $N(t-)$, can improve precision of a treatment effect without appreciable bias. 
The bloodstream infection application illustrates the use of the method in data with substantial heterogeneity between patients in both recurrence patterns and survival.

The low-action simulations identify an important practical limitation. If recurrent events are sparse, the observed history contains little information about subject-specific heterogeneity. In this setting, the benefit of augmentation was small or absent,
and the most complex prediction model could worsen coverage. Thus, the complexity of the augmentation model should be matched to the amount of
information available in the event histories, rather than being increased automatically.

The proposed method is available in the {\tt R} package {\tt mets} and illustrated in an accompanying vignette \citep{metsR}. Its computing time grows approximately linearly with sample size, making it practical for large datasets. 

The same augmentation theory may also be applicable to marginal regression models for composite endpoints combining recurrent and terminal events \citep{mao-lin-2016}. This would require making a slight modification to the risk set to reflect the chosen endpoint, but the IPCW representation and the
censoring-augmentation argument can be applied in the same way. This extension would be particularly useful for studies in which several clinically relevant recurrent outcomes are combined.
Moreover, future investigations might be devoted to improving performance of the augmented estimators in efficiency gain when events are sparse, 
and to clarifying the robustness properties, e.g., by the identification of general conditions under which the augmentation procedure fully corrects bias from a misspecified censoring model.

%
%
\section*{Data availability}
Data used in the application section were provided under license for this study and are available from the authors with permission from the University Hospital of Copenhagen.

\bibliography{ghosh-lin}

@Book{		  abgk-book,
  title		= {Statistical Models Based on Counting Processes},
  author	= {Andersen, P. K. and Borgan, {\O}. and Gill, R. D. and
		  Keiding, N.},
  publisher	= {Springer},
  year		= {1993},
  address	= {New York}
}

@Article{stru:kalb:1986,
  Title                    = {Misspecified Proportional Hazard Models},
  Author                   = {Struthers, C. A. and Kalbfleisch, J. D.},
  Journal                  = {Biometrika},
  Year                     = {1986},
  Pages                    = {363--369},
  Volume                   = {73}
}

@article{Scheike2026,
  author  = {Scheike, Thomas H.},
  title   = {Two-stage recurrent events random effects models},
  journal = {Lifetime Data Analysis},
  year    = {2026},
  volume  = {32},
  number  = {1},
  pages   = {6},
  doi     = {10.1007/s10985-025-09680-z},
  url     = {https://doi.org/10.1007/s10985-025-09680-z}
}

@Article{Blanche2022,
  author          = {Blanche, Paul Frédéric and Holt, Anders and Scheike, Thomas},
  journal         = {Lifetime data analysis},
  title           = {On logistic regression with right censored data, with or without competing risks, and its use for estimating treatment effects.},
  year            = {2023},
  volume       = {29},
  pages        = {441--482}
}

@Article{	  ag-82,
  title		= {{C}ox's regression model for counting processes: A large
		  sample study},
  author	= {Andersen, P. K. and Gill, R. D.},
  journal	= {Ann. Statist.},
  year		= {1982},
  pages		= {1100-1120},
  volume	= {10}
}

@Article{	  bang:tsiatis:2000,
  title		= {Estimating medical costs with censored data},
  author	= {Bang, H. and Tsiatis, A.~A.},
  journal	= {Biometrika},
  year		= {2000},
  pages		= {329--343},
  volume	= {87}
}

@Book{		  cook-lawless-2007,
  title		= {The statistical analysis of recurrent events},
  author	= {Cook, Richard J and Lawless, Jerald},
  publisher	= {Springer Science \& Business Media},
  year		= {2007}
}

@Article{	  cook-lawless-1997,
  title		= {Marginal analysis of recurrent events and a terminating
		  event},
  author	= {Cook, Richard J and Lawless, Jerald F},
  journal	= {Statistics in medicine},
  year		= {1997},
  number	= {8},
  pages		= {911--924},
  volume	= {16},
  publisher	= {Wiley Online Library}
}

@Article{	  ghosh2002,
  title		= {Marginal regression models for recurrent and terminal
		  events},
  author	= {Ghosh, Debashis and Lin, DY},
  journal	= {Statistica Sinica},
  year		= {2002},
  pages		= {663--688},
  publisher	= {JSTOR}
}

@Article{	  ghosh-lin-2000,
  title		= {Nonparametric analysis of recurrent events and death},
  author	= {Ghosh, Debashis and Lin, DY},
  journal	= {Biometrics},
  year		= {2000},
  number	= {2},
  pages		= {554--562},
  volume	= {56},
  publisher	= {Wiley Online Library}
}

@Article{	  marso,
  title		= {Liraglutide and cardiovascular outcomes in type 2 diabetes},
  author	= {Marso, S.P. and Daniels, G.H. and Brown-Frandsen, K. et al.},
  journal	= {N Engl J Med},
  year		= {2016},
  pages		= {311--322},
  volume	= {375},
  publisher	= {Wiley Online Library}
}

@Manual{	  metsr,
  title		= {mets: Analysis of Multivariate Event Times},
  author	= {Klaus K. Holst and Thomas Scheike},
  note		= {R package version 1.3.13},
  year		= {2026},
  url		= {http://lava.r-forge.r-project.org/}
}

@Article{	  lin:wei:yang:ying:2000,
  title		= {Semiparametric regression for the mean and rate functions
		  of recurrent events},
  author	= {Lin, D. Y. and Wei, L.J. and Yang, I. and Ying, Z.},
  journal	= {J. Roy. Stat. Soc. Ser. B},
  year		= {2000},
  pages		= {711--730},
  volume	= {62}
}

@InProceedings{	  robins-rotnitzky:1992,
  title		= {Recovery of information and adjustment of dependent
		  censoring using surrogate markers},
  author	= {Robins, J.~M. and Rotnitzky, A.},
  booktitle	= {AIDS Epidemiology-Methodological Issues},
  year		= {1992},
  pages		= {24-33},
  publisher	= {Birkh\"{a}user, Boston},
  editors	= {Jewell, N. and Dietz, K. and Farewell, V.}
}

@Article{	  tribler2018,
  title		= {Catheter-related bloodstream infections in patients with
		  intestinal failure receiving home parenteral support: risks
		  related to a catheter-salvage strategy},
  author	= {Siri Tribler and Brandt, {Christopher F.} and Fuglsang,
		  {Kristian A.} and Michael Staun and Per Broebech and Moser,
		  {Claus E.} and Thomas Scheike and Jeppesen, {Palle B.}},
  journal	= {American Journal of Clinical Nutrition},
  year		= {2018},
  number	= {5},
  pages		= {743--753},
  volume	= {107},
  doi		= {10.1093/ajcn/nqy010},
  issn		= {0002-9165},
  language	= {English},
  publisher	= {American Society for Nutrition}
}

@Book{		  tsiatis2006semiparametric,
  title		= {Semiparametric theory and missing data},
  author	= {Tsiatis, A.A.},
  publisher	= {Springer Verlag},
  year		= {2006}
}

@Book{		  van2003unified,
  title		= {Unified methods for censored longitudinal data and
		  causality},
  author	= {Van der Laan, M.J. and Robins, J.M.},
  publisher	= {Springer Verlag},
  year		= {2003}
}

@Article{	  wei:lin:weissfeld:1989,
  title		= {Regression analysis of multivariate incomplete failure
		  time data by modeling marginal distributions},
  author	= {Wei, L. J. and Lin, D.Y. and Weissfeld, L.},
  journal	= {J. Amer. Statist. Assoc.},
  year		= {1989},
  pages		= {1065--1073},
  volume	= {84}
}

@Article{	  akacha2018request,
  author	= {Akacha, M and Binkowitz, B and Bretz, F and Fritsch, A and Hougaard, P and others},
  journal	= {Recurrent Event Qualification Opinion Consortium},
  title		= {Request for CHMP Qualification Opinion: Clinically
		  Interpretable Treatment Effect Measures based on Recurrent
		  Event Endpoints that Allow for Efficient Statistical
		  analyses},
  year		= {2018}
}

@Article{	  mao-lin-2016,
  author	= {Mao, Lu and Lin, D. Y.},
  journal	= {Biostatistics},
  title		= {{Semiparametric regression for the weighted composite
		  endpoint of recurrent and terminal events}},
  year		= {2016},
  issn		= {1465-4644},
  month		= {12},
  number	= {2},
  pages		= {390-403},
  volume	= {17},
  doi		= {10.1093/biostatistics/kxv050},
  eprint	= {https://academic.oup.com/biostatistics/article-pdf/17/2/390/6692920/kxv050.pdf},
  url		= {https://doi.org/10.1093/biostatistics/kxv050}
}

@Book{		  cook-lawless-2018,
  author	= {Cook, Richard J and Lawless, Jerald F},
  publisher	= {CRC Press},
  title		= {Multistate models for the analysis of life history data},
  year		= {2018}
}

@Book{		  andersen-ravn-2023,
  author	= {Andersen, Per Kragh and Ravn, Henrik},
  publisher	= {CRC Press},
  title		= {Models for Multi-State Survival Data: Rates, Risks, and
		  Pseudo-Values},
  year		= {2023}
}

@Article{	  furberg-leader-2022,
  author	= {Furberg, Julie Kjærulff and Rasmussen, Søren and
		  Andersen, Per Kragh and Ravn, Henrik},
  journal	= {Pharmaceutical Statistics},
  title		= {Methodological challenges in the analysis of recurrent
		  events for randomised controlled trials with application to
		  cardiovascular events in LEADER},
  year		= {2022},
  number	= {1},
  pages		= {241-267},
  volume	= {21},
  doi		= {https://doi.org/10.1002/pst.2167},
  eprint	= {https://onlinelibrary.wiley.com/doi/pdf/10.1002/pst.2167},
  url		= {https://onlinelibrary.wiley.com/doi/abs/10.1002/pst.2167}
}

@Article{	  cortese-scheike-2022,
  author	= {Cortese, Giuliana and Scheike, Thomas H.},
  journal	= {Journal of the Royal Statistical Society Series C},
  title		= {Efficient estimation of the marginal mean of recurrent
		  events},
  year		= {2022},
  number	= {5},
  pages		= {1787-1821},
  volume	= {71}
}

@Article{	  pwp-1981,
  author	= {Prentice, Ross L and Williams, Benjamin J and Peterson,
		  Arthur V},
  journal	= {Biometrika},
  title		= {On the regression analysis of multivariate failure time
		  data},
  year		= {1981},
  number	= {2},
  pages		= {373--379},
  volume	= {68},
  publisher	= {Oxford University Press}
}

\bibliographystyle{apalike}

\section*{Appendix}

\subsection{Appendix A: Proof of the estimating equation decomposition}
 
With algebraic manipulations and using the uniform convergence of $\hat G(t)$ to $G(t) $, we can write the estimating equation for $\beta$  in the form given in  \eqref{eq1:phiR} and \eqref{eq1:phiC}:
 \begin{align*}
n^{-1/2}  U(\beta)  & =   n^{-1/2}  \sum_{i=1}^n \int_0^{\tau} \left( X_i - \bar X(\beta, t) \right) w_i(t) dM_i(t,X)    \nonumber \\
 &  + n^{-1/2}  \sum_{i=1}^n \int_0^{\tau} \left( X_i - \bar X(\beta, t) \right) 
\left( \frac{\hat G(t)}{\hat G(T_i \wedge t)} - \frac{ G(t)}{ G(T_i \wedge t)} \right)
I(C_i \geq D_i \wedge t)  dM_i(t,X).
\end{align*}
Then, using the martingale representation of the K-M estimator,  we obtain
 \begin{align*}
n^{1/2}  \left( \frac{\hat G(t)}{\hat G(T_i \wedge t)} - \frac{ G(t)}{ G(T_i \wedge t)} \right) & =    - n^{-1/2} G(t) \frac{ I(T_i < t ) }{ G(T_i )} \sum_{j=1}^n \int_{T_i}^t  \frac{1}{Y_{\bullet}(s)} dM_j^C(s)  + o_P(1) \\
   & =    - n^{-1/2} G(t) \frac{ I(T_i < t ) }{ G(T_i )} \sum_{j=1}^n \int_{T_i}^t  \frac{1}{G(s) S(s)} dM_j^C(s) + o_P(1), 
 \end{align*}
 with $ Y_{\bullet}(s) = \sum_{i=1}^n Y_i(s)$. Because of the identity $ Y_{\bullet}(s) = n \hat G(s) \hat S(s)$, the last equation is obtained using the consistency of $\hat G(t)$ and $\hat S(t)$.
Then, plugging this identity into the above estimating equation  $n^{-1/2}  U(\beta) $ and changing the order of integration, we observe that
 \begin{align*}
  & n^{-1/2}  \sum_{i=1}^n \int_0^{\tau} \left( X_i - \bar X(\beta, t) \right) 
\left( \frac{\hat G(t)}{\hat G(T_i \wedge t)} - \frac{ G(t)}{ G(T_i \wedge t)} \right)
I(C_i \geq D_i \wedge t)  dM_i(t,X) \\
 & = n^{-1/2}  \sum_{i=1}^n \int_0^{\tau} \left( X_i - \bar X(\beta, t) \right) 
I(T_i < t) w_i(t)  
\left( - n^{-1}  \sum_{j=1}^n \int_{T_i}^t \frac{1}{G(s) S(s)} dM_j^C(s)  \right)
  dM_i(t,X) \\
  & = n^{-1/2}  \sum_{i=1}^n \int_0^{\tau} 
   \left( - n^{-1}  \sum_{j=1}^n \int_{t}^{\tau}  \left( X_j - \bar X(\beta, s) \right)  
   I(T_j < t )  w_j(s) dM_j(s,X)    \right)
    \frac{1}{G(t) S(t)} dM_j^C(t) \\
     & = n^{-1/2}  \sum_{i=1}^n \int_0^{\tau} \hat q(t)  \frac{1}{G(t) S(t)} dM_i^C(t)
\end{align*}
with
$   \hat q(t)   = - n^{-1}  \sum_{i=1}^n \int_{t}^{\tau}  \left( X_i - \bar X(\beta, s) \right)  
   I(T_i < t)  w_i(s) dM_i(s,X) 
$. \\
Replacing  $\bar X(\beta, t)$ with the limit $ \bar x(\beta, t) = s_1(\beta, t) / s_0(\beta, t) $, where 
$s_k(\beta, t) = E(G(t) X^{\otimes k} \exp{(\beta^T X_i)} )$,
and, by the martingale central limit theorem,  replacing $\hat q(t)$ with its limit
$$
 q(t) = - E [ \int_t^{\tau}  \{X_i - \bar x(\beta, s) \} \ I(T_i < t)  w_i(s) dM_i(s,X) ] , 
$$
we obtain the estimating equation based on an influence function decomposed in 
$\phi_i^R(X,\beta_0)$ and $\phi_i^C(X,\beta_0)$, as given in \eqref{eq1:phiR} and \eqref{eq1:phiC}.

\subsection{Appendix B: Robustness under the null hypothesis in the RCT setting}

Let us consider the estimating equation  of the Ghosh-Lin estimator given in \eqref{eq:robGL}. 
Let $\beta^*=(\beta_1^*,\gamma^*)$ be the solution to the asymptotic Ghosh-Lin estimating equation, 
where the parameter that represents the treatment effect is denoted by $\beta^*_1$.
Let us then simply compute the $s_j(t)$ and $s_j(t,\beta)$ for $j=0,1$ with $\beta_1^*=0$, and observe that this solves the estimating equation.

Let $s_{1,1}(\cdot)$ denote the component of $s_1(\cdot)$ related to the covariate $A$.
First, consider condition (1),  using a correctly specified censoring model and stabilized weights on the form 
$w(t) = G(t, A) I(C > D \wedge t)/G(D \wedge t; A, X)$ for fitting the Ghosh-Lin model. 
Let us assume $P(A=1)=\pi_1$. Then, let us set $\beta^*_1=0$ and compute the different components of the estimating equation that it solves. 
Assuming the correctly specified censoring model, we get 
\begin{align*}
s_{1,1}(t) & = E[ w(t) A \, \lambda(t,X) ]  & =  & \, \pi_1 G(t;1) E[ \lambda(t,X) ] \\
s_{0}(t) & = E[ w(t) \lambda(t,X) ]      & =  & \, [ \pi_1 G(t;1) + (1- \pi_1)G(t;0)] \, E[ \lambda(t,X)] \\
s_{1,1}(t,\beta) & = E[ w(t) A \exp(X^T \gamma^* )] &  =  & \, \pi_1 G(t;1) E[ \exp(X^T \gamma^*) ] \\
s_{0}(t,\beta) & = E[ w(t)  \exp(X^T \gamma^*) ]  &  = & \, [ \pi_1 G(t;1) + (1- \pi_1)G(t;0) ] \, E[ \exp(X^T \gamma^*) ]  
\end{align*}
and therefore, we obtain that $s_{1,1}(t) - \frac{s_{1,1}(t,\beta)}{s_0(t,\beta)} s_0(t) = 0 $.

Let us now assume that condition (2) is verified. We thus consider an additive hazard model stratified on $A$ for the censoring variable, and a working model $G^*$ used 
for constructing the hazards weights, 
$w^*(t,A)= G^*(t;A)  I(C > D \wedge t)/G^*(D \wedge t; A)$ in the  Ghosh-Lin estimating equation. 
Taking arguments  similar to \cite{Blanche2022} (see, subsection 7.2.1), we can show that the working model $G^*$ leads to an hazard function on the form
$\lambda_0^*(t)+ A \lambda_1(t)$. Then,
we get 
\begin{align}
E( w^*(t,A) | A,X) & = G^*(t;A) E\left\{ \frac{G(D \wedge t; A,X)  }{G^*(D \wedge t;A)} \mid X,A\right\} \nonumber \\
                   & =  G^*(t,A) \, E[ \nu(t,X,D \wedge t) | X] 
\label{p-form}.
\end{align}
Therefore, we compute 
\begin{align*}
s_{1,1}(t) & = E[ w(t) A \lambda(t,X) ]  & =  &\, \pi_1 G^*(t;1) E[ \nu(t,X) \lambda(t,X) ] \\
s_{0}(t) & = E[ w(t) \lambda(t,X) ]      & =  & \, [ \pi_1 G^*(t;1) + (1- \pi_1)G^*(t;0)] \, E[ \nu(t,X) \lambda(t,X)] \\
s_{1,1}(t,\beta) & = E[ w(t) A \exp(X^T \gamma^* )] &  =  & \, \pi_1 G^*(t;1) E[ \nu(t,X) \exp(X^T \gamma^*)  ] \\
s_{0}(t,\beta) & = E[ w(t)  \exp(X^T \gamma^*) ]  &  = & \, [ \pi_1 G^*(t;1) + (1- \pi_1)G^*(t;0)] \, E[ \nu(t,X) \exp(X^T \gamma^*) ],
\end{align*}
and again we obtain that $s_{1,1}(t) - \frac{s_{1,1}(t,\beta)}{s_0(t,\beta)} s_0(t) = 0 $.

If we use a censoring working model that stratifies according to both $A$ and strata based on $X$ (e.g., $\tilde X$) 
the above arguments are still valid because equation \eqref{p-form} remains valid.

\end{document}